\documentclass[useAMS,usenatbib]{mn2e}

\usepackage{graphicx}
\usepackage{graphics}
\usepackage{natbib}
\usepackage{amsmath}
\usepackage{subfig}

\pdfoutput=1

\title[Optical Polarimetry of the Inner Crab Nebula and Pulsar]
  {Optical Polarimetry of the Inner Crab Nebula and Pulsar}
\author[P. Moran et al.]
  {P.~Moran,$^1$
  A.~Shearer,$^1$\thanks{E-mail: andy.shearer@nuigalway.ie} R. P.~Mignani,$^{2,3,4}$ A.~S\l{}owikowska,$^4$
  A.~De Luca,$^{3,5}$ 
  \newauthor C.~Gouiff\`es,$^{6,7}$ and P.~Laurent$^{6,8}$\\
  $^1$Centre for Astronomy, NUI Galway, Ireland\\
  $^2$MSSL UCL, Holmbury St. Mary, Dorking, Surrey, RH5 6NT, UK\\
  $^3$INAF - Istituto di Astrofisica Spaziale e Fisica Cosmica Milano, via E. Bassini 15, 20133, Milano, Italy\\
  $^4$Kepler Institute of Astronomy, University of Zielona G\'ora, Lubuska 2, 65-265, Zielona G\'ora, Poland\\
  $^5$INFN – Istituto Nazionale di Fisica Nucleare, sezione di Pavia, via A. Bassi 6, 27100 Pavia, Italy\\
  $^6$CEA, IRFU, Service d'Astrophysique, 91191 Gif-sur-Yvette, France\\
  $^7$AIM, CEA/CNRS/Universit\'e Paris Diderot, SAp, Saclay, 91191 Gif-sur-Yvette, France\\
  $^8$APC, 10 rue Alice Domont et Leonie Duquet, 75205 Paris Cedex 13, France\\
  }
\date{Released 2012 Xxxxx XX}

\pagerange{\pageref{firstpage}--\pageref{lastpage}} \pubyear{2012}

\def\LaTeX{L\kern-.36em\raise.3ex\hbox{a}\kern-.15em
    T\kern-.1667em\lower.7ex\hbox{E}\kern-.125emX}

\begin{document}

\label{firstpage}

\maketitle


\begin{abstract}
Time-resolved polarisation measurements of pulsars offer an unique insight into the geometry of their emission regions. Such measurements provide observational constraints on the different models proposed for the pulsar emission mechanisms. Optical polarisation data of the Crab Nebula was obtained from the HST archive. The dataset consists of a series of observations of the nebula taken with the HST/ACS. We produced polarisation vector maps of the inner nebula and measured, for the first time, the degree of linear polarisation (P.D.) and the position angle (P.A.) of the pulsar's integrated pulse beam, and of its nearby synchrotron knot. This yielded $\rm P.D.=5.2\pm0.3\%$ and $\rm P.A.=105.1\pm1.6\degr$ for the pulsar, and $\rm P.D.=59.0\pm1.9\%$ and $\rm P.A.=124.7\pm1.0\degr$ for the synchrotron knot. This is the first high-spatial resolution multi-epoch study of the polarisation of the inner nebula and pulsar. None of the main features in the nebula show evidence of significant polarisation evolution in the period covered by these observations. The results for the pulsar are consistent with those obtained by \citet{Aga} using the high-time resolution photo-polarimeter OPTIMA, once the DC component has been subtracted. Our results clearly prove that the knot is the main source of the DC component.
\end{abstract}




\section{Introduction}

Strong polarisation is expected when the pulsar optical emission is generated by synchrotron radiation. \citet{Shklovsky} suggested that the continuous optical radiation from the Crab Nebula was due to synchrotron radiation. This was later confirmed by \citet{Dombrovsky} and \citet{Vashakidze} who found that the optical radiation was polarised. Incoherent synchrotron emission follows a simple relationship between its polarisation profile and underlying geometry. Hence, optical polarisation measurements of pulsars provide an unique insight into the geometry of their emission regions, and therefore observational constraints on the theoretical models of the emission mechanisms. From an understanding of the emission geometry, one can limit the competing models for pulsar emission, and hence understand how pulsars work - a problem which has eluded astronomers for almost 50 years.

Polarimeters are sensitive in the optical, but the majority of pulsars are very faint at these wavelengths with V $\ge$ 25 \citep{Shearer}. Polarimetry in the very high-energy domain, X-ray and gamma-ray, using instruments on board space telescopes, is of limited sensitivity. So far, detailed results have only been reported for the Crab pulsar \citep{Weisskopf,Dean,Forot}. Although the number of pulsars detected in the optical is growing, only five pulsars have had their optical polarisation measured; Crab \citep{Wampler, Kristian, Smith, Aga}, Vela \citep{Wagner, Mignani}, PSR B0540-69 \citep{Middleditch,Chanan,Wagner,Mignani10} PSR B0656+14 \citep{Kern}, and PSR B1509-58 \citep{Wagner}. Nonetheless, the optical currently remains invaluable for polarimetry in the energy domain above radio photon energies. The Crab pulsar, being the brightest optical pulsar  with V $\approx 16.8$ \citep{Nasuti}, has had several measurements of its optical polarisation, including both phase-averaged and phase-resolved studies.

The first phase-resolved observations of the optical linear polarisation of the Crab pulsar were those of \citet{Wampler}, \citet{Cocke}, and \citet{Kristian}. Those studies showed that the polarisation position angle changes through each peak in the pulsar lightcurve, and that the degree of polarisation falls and rises within each peak, reaching its minimum value shortly after the pulse peak. These observations were limited to the main and inter pulse phase regions only, because at the time it was thought that the pulsar radiated its optical emission through the pulse peaks only. However, a number of phase-resolved imaging observations of the pulsar \citep{Peterson,Jones,Percival,Golden} showed that the optical emission actually persists throughout the pulsar's entire rotation cycle, at the level of $\sim1\%$ of the maximum main-pulse intensity.

With this in mind, observations of the optical linear polarisation of the Crab pulsar were made by \citet{Jones} and \citet{Smith}. These results confirmed the previous observations, and were the first studies to reveal the polarisation profile of the pulsar during the bridge and off-pulse phase regions. They also found that the off-pulse region is highly polarised. The degree of polarisation was 70\% and $47\pm10\%$ for \citet{Jones} and \citet{Smith}, respectively. \citet{Aga} report the most detailed phase-resolved observations of the optical linear polarisation of the Crab pulsar. Their results are consistent with previous observations albeit with better definition and statistics, and can be explained in the context of the two-pole caustic model \citep{Dyks}, the outer-gap model \citep{Romani,Takata}, and the striped-wind model \citep{Petri}.

Detailed observations of the inner nebula, in the optical and X-rays, have revealed a torus structure, that is bisected by oppositely-directed jets, knots, and a series of highly variable synchrotron wisps \citep{Scargle,Hester,Weisskopf2000}. A bright knot of synchrotron emission is located 0\farcs65 SE of the pulsar. It is highly polarised and is slightly variable in both its location and its brightness \citep{Hester,Hester02,Hester08}. \citet{Komissarov} proposed that it is radiation from an oblique termination shock in the pulsar wind nebula. In this model, the Earth line-of-sight is tangent to the flow at the position of the knot, hence the intensity is  Doppler boosted for synchrotron emission in the mildly relativistic post-shock flow. The results of their relativistic MHD simulations show that the knot is highly variable and can dominate the gamma-ray synchrotron emission. It has been suggested that this knot is responsible for the highly polarised off-pulse emission seen in time-resolved observations in the optical \citep{Aga,Aga13} and gamma-rays \citep{Forot}. The knot has no counterpart to the NW of the pulsar, but there is a faint second knot located 3\farcs8 SE of the pulsar, the so called outer knot. 

The wisps consist of Wisp 1, a Thin wisp located 1\farcs8 NW of the pulsar, and the Counter wisp located 8\farcs3 SE of the pulsar. Wisp 1, located 7\farcs3 NW of the pulsar, breaks into three separate and distinct components; 1-A, 1-B, and 1-C \citep{Scargle}. However, one must note that this does not represent the constant configuration of the wisps. The wisps are interpreted as magnetic flux tubes that undergo unstable synchrotron cooling \citep{Hester98}. A number of follow-up observations of the inner nebula have confirmed the presence of the wisps \citep{Hester,Bietenholz}. The NW wisps are more prominent than the SE wisps due to Doppler beaming of the flow. \citet{Schweizer} studied the behaviour of the wisps NW of the pulsar in both the optical and X-rays. They observed that the wisps form and move off from the region associated with the termination shock of the pulsar wind, roughly once per year. Moreover, they found that the precise locations of the NW wisps in the optical and X-rays are similar but not exactly coincident, with X-ray wisps located closer to the pulsar. This would suggest that the optical and X-ray wisps are not produced by the same particle distribution. In terms of MHD models, they found that the optical wisps are more strongly Doppler-boosted than the X-ray wisps. For a more detailed review of the Crab Nebula see \citet{Hester08}. \\
\indent The first optical linear polarisation maps of the Crab Nebula were produced by \citet{Oort}, \citet{Hiltner}, and \citet{Woltjer}. X-ray observations of the linear polarisation of the nebula, in the range 2.6--5.2 keV, yield polarisation of 19\% at a position angle of 152--156\degr within a radius of 3\degr of the pulsar \citep{Weisskopf}. These results are in agreement with the optical measurements of the polarisation, which give polarisation of 19\% at a position angle 162\degr for the central nebular region within a radius of $\approx$ 0\farcm5 \citep{Oort}. \citet{Dean} measured the polarisation of the Crab Nebula and pulsar in the off-pulse phase using the INTEGRAL/SPI telescope, and showed that the polarisation E-vector (124$\pm11\degr$) is aligned with the spin-axis of the neutron star (\citet{Kaplan08}; 110$\pm2\pm9\degr$, where the first uncertainty is the measurement uncertainty and the second is from the reference frame uncertainty). This result was later confirmed by \citet{Forot}, in the off-pulse phase region using the INTEGRAL/IBIS telescope (120.6$\pm8.5\degr$), and has also been seen in optical observations \citep{Smith,Aga}. The SPI and IBIS measurements both encompass the entire nebula and pulsar, and so are dominated by nebular emission. As with the optical observations \citep{Smith,Aga}, they found the off-pulse region to be highly polarised. \\ 
\indent The purpose of this work is two fold. Firstly, we want to check the polarisation of the pulsar, knot, and wisps for variability. It is difficult to determine the polarisation for objects embedded in a strong nebular background. So, in order to determine the Crab pulsar's polarisation profile, we need to know the level of background polarisation. Therefore, the second purpose of this work is to accurately map the polarisation of the inner Crab Nebula. This will then act as a guideline for future time-resolved polarisation measurements of the Crab pulsar using the Galway Astronomical Stokes Polarimeter (GASP). This is an ultra-high-speed, full Stokes, astronomical imaging polarimeter based on the Division of Amplitude Polarimeter (DOAP). It has been designed to resolve extremely rapid variations in objects such as optical pulsars and magnetic cataclysmic variables \citep{Kyne}.


\section{Observations and Analysis}

\begin{table*}
 \caption{Summary of the HST/ACS observations of the Crab Nebula. The filters used were F606W ($\lambda=590.70$ nm, $\Delta\lambda=250.00$ nm) and F550M ($\lambda=558.15$ nm, $\Delta\lambda=54.70$ nm).}
 \begin{tabular}{|c|c|c|l|c|c|}
  \hline
  Date & Exposure (s) & Filter & Polariser & Roll-Angle (PA\_V3) (\degr) & Pulsar Position on Chip (x,y) \\
  \hline
  2003 Aug 08  &2$\times$1200  &F606W  &POL0V  &87.6 & 1320.30 1045.02\\
  &2$\times$1200  &  &POL60V  & &\\
  &2$\times$1200  &  &POL120V  & &\\
  \hline
  2005 Sep 06 &2$\times$1150  &F550M  &CLEAR2L &87.2 & 1316.23 1034.54\\
  &2$\times$1150  &F606W  &POL0V & & \\
  &2$\times$1150  &  &POL60V & & \\
  &2$\times$1150  &  &POL120V &  &\\
  \hline
  2005 Sep 15 &2$\times$1150  &F550M  &CLEAR2L &87.4 & 1315.58 1042.01\\
  &2$\times$1150  &F606W  &POL0V & & \\
  &2$\times$1150  &  &POL60V & & \\
  &2$\times$1150  &  &POL120V & &\\
  \hline
  2005 Sep 25 &2$\times$1150  &F550M  &CLEAR2L  &87.6 & 1315.92 1041.17\\
  &2$\times$1150  &F606W  &POL0V & & \\
  &2$\times$1150  &  &POL60V & &\\
  &2$\times$1150  &  &POL120V & &\\
  \hline
  2005 Oct 02 &2$\times$1150  &F550M  &CLEAR2L &87.8 & 1315.97 1040.11\\
  &2$\times$1150  &F606W  &POL0V & & \\
  &2$\times$1150  &  &POL60V & &\\
  &2$\times$1150  &  &POL120V & &\\
  \hline
  2005 Oct 12 &2$\times$975   &F550M  &CLEAR2L  &88.0 & 1316.44 1039.62\\
  &2$\times$975   &F606W  &POL0V & &\\
  &2$\times$1000  &  &POL60V & &\\
  &2$\times$1000  &  &POL120V & &\\
  \hline
  2005 Oct 22 &2$\times$1150  &F550M  &CLEAR2L &88.3 & 1316.66 1038.13\\
  &2$\times$1150  &F606W  &POL0V & & \\
  &2$\times$1150  &  &POL60V & &\\
  &2$\times$1150  &  &POL120V & &\\
  \hline
  2005 Oct 30 &2$\times$1150  &F550M  &CLEAR2L &88.6 & 1316.40 1036.31\\
  &2$\times$1150  &F606W  &POL0V & & \\
  &2$\times$1150  &  &POL60V  & &\\
  &2$\times$1150  &  &POL120V & &\\
  \hline
  2005 Nov 08 &2$\times$1150  &F550M  &CLEAR2L &89.0 & 1316.23 1034.54\\
  &2$\times$1140  &F606W  &POL0V & & \\
  &2$\times$1150  &  &POL60V  & &\\
  &2$\times$1150  &  &POL120V & &\\
  \hline
  2005 Nov 16 &2$\times$1150  &F550M  &CLEAR2L  &89.6 & 1316.57 1031.40\\
  &2$\times$1150  &F606W  &POL0V & & \\
  &2$\times$1150  & &POL60V  & &\\
  &2$\times$1150  & &POL120V  & &\\
  \hline
  2005 Nov 25 &2$\times$1150  &F550M  &CLEAR2L  &90.6 & 1315.74 1025.14\\
   &2$\times$1150  &F606W  &POL0V & & \\
   &2$\times$1150  &  &POL60V  & &\\
   &2$\times$1150  &  &POL120V & &\\
  \hline
  2005 Dec 05 &2$\times$1150  &F550M  &CLEAR2L &120.0 & 1279.52 871.13\\
  &2$\times$1150  &F606W  &POL0V & & \\
  &2$\times$1150  &  &POL60V  & &\\
  &2$\times$1150  &  &POL120V  & &\\
  \hline
  2005 Dec 14 &2$\times$1150  &F550M  &CLEAR2L &125.0 & 1267.44 851.24\\
  &2$\times$1150  &F606W  &POL0V & & \\
  &2$\times$1150  &  &POL60V & &\\
  &2$\times$1150  &  &POL120V & &\\
  \hline
 \end{tabular}
\end{table*}

\begin{figure*}
\includegraphics[width=100mm]{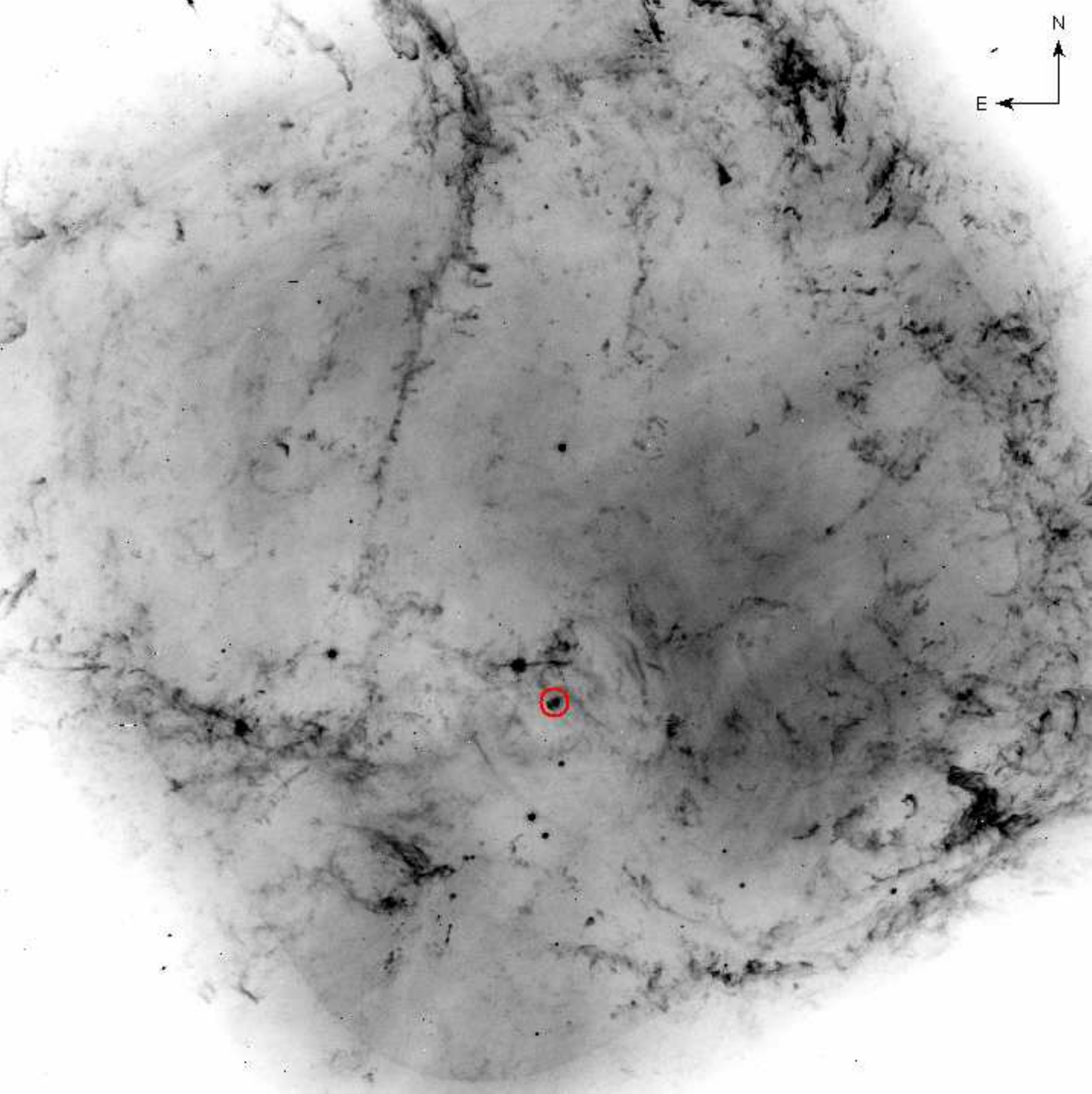}
\caption{HST/ACS image of the inner Crab Nebula (2005 Sep 06, FOV $\approx102\arcsec\times102\arcsec$, F606W, POL0V). The location of the pulsar and inner knot is marked by the circle.}
\label{figure1}
\end{figure*}

\begin{figure*}
\includegraphics[width=100mm]{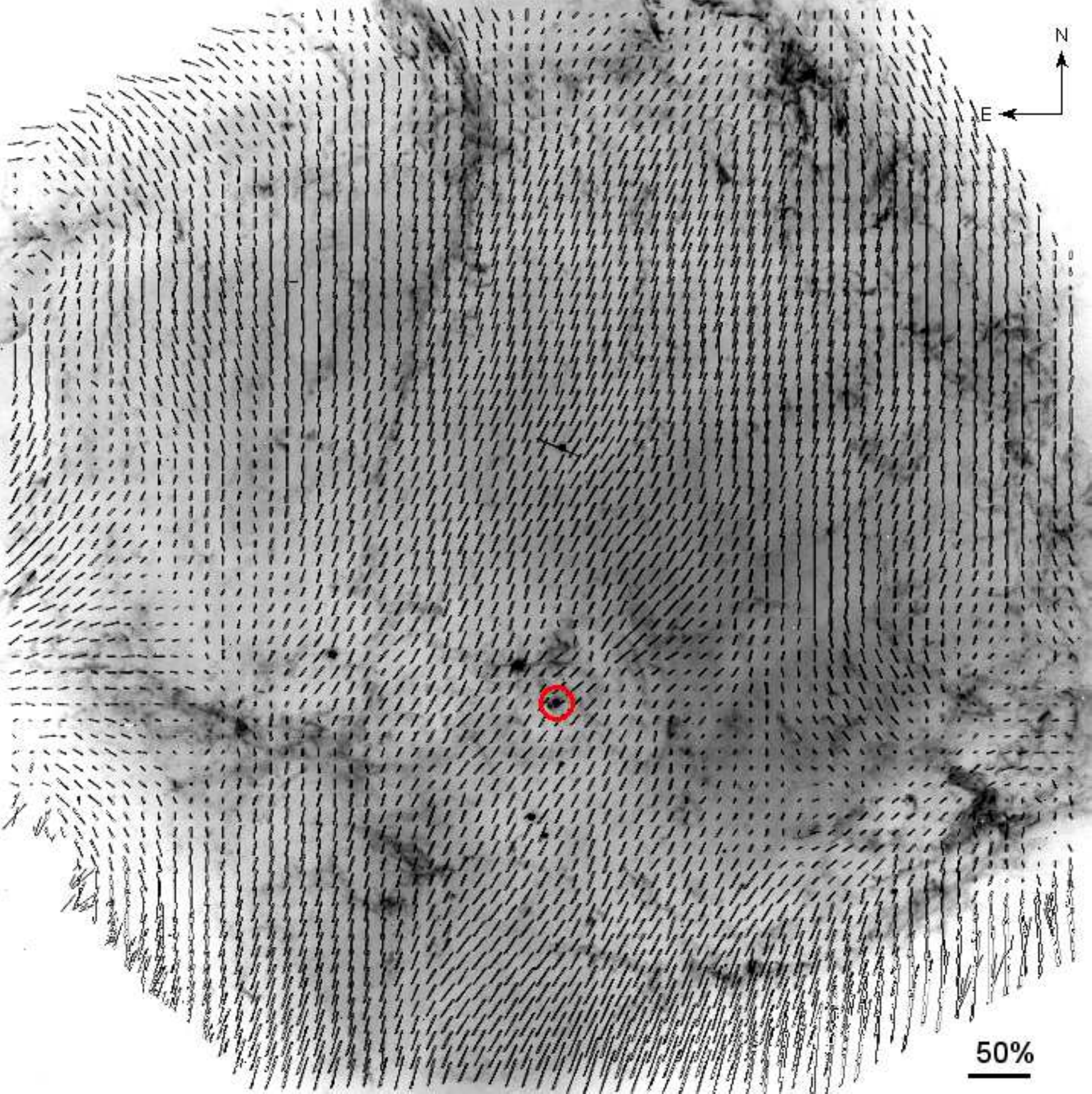}
\caption{Polarisation vector map of the inner Crab Nebula superimposed on the nebula (2005 Sep 06, FOV $\approx102\arcsec\times102\arcsec$). The location of the pulsar and inner knot is marked by the circle. The legend shows the vector magnitude for 50\% polarisation.}
\label{figure2}
\end{figure*}

\begin{figure*}
\includegraphics[width=95mm]{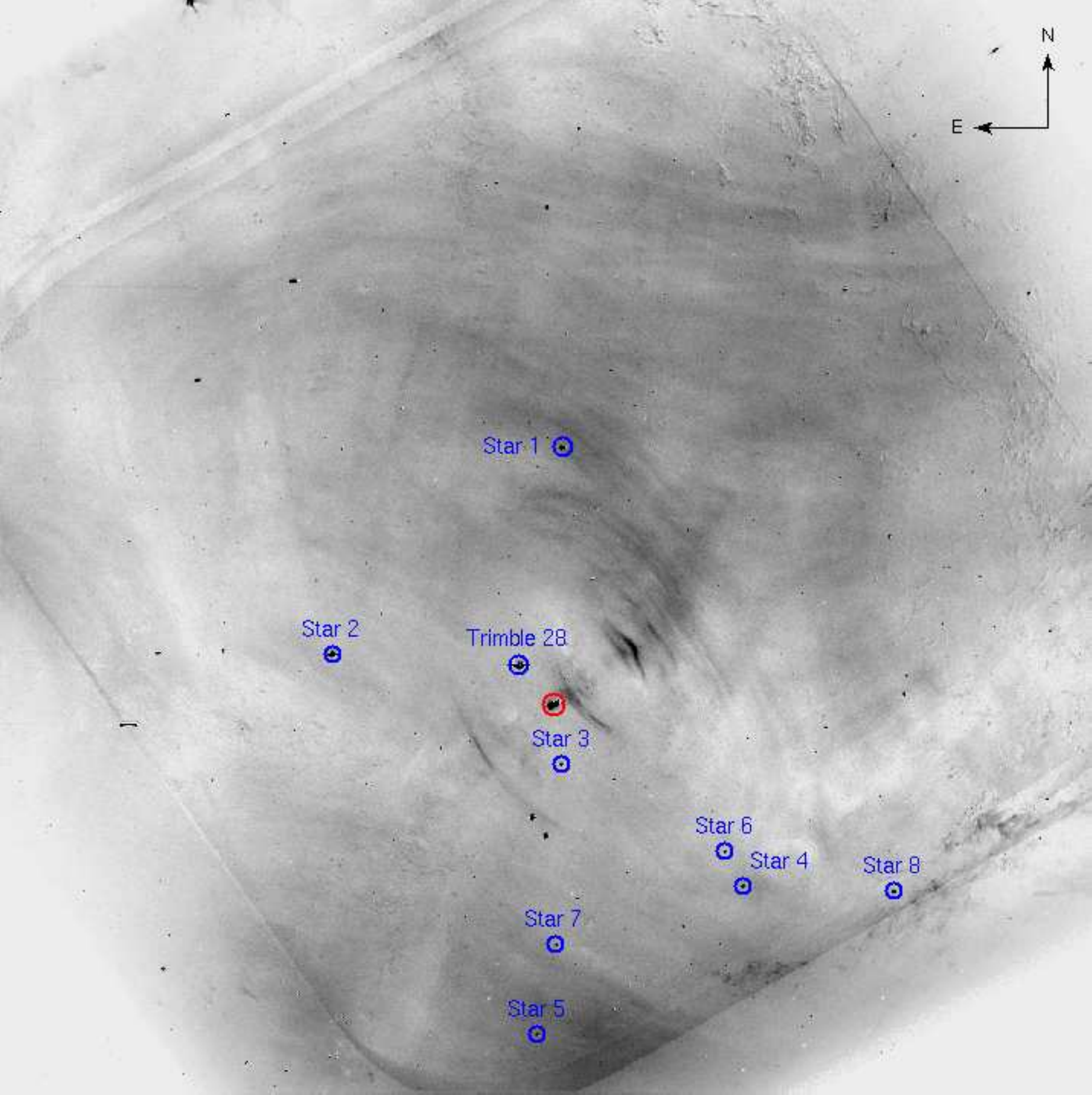}
\caption{Polarised flux map of the inner Crab Nebula (2005 Sep 06, FOV $\approx102\arcsec\times102\arcsec$) with the stars for analysis marked. The location of the pulsar and inner knot is marked by the circle.}
\label{figure3}
\end{figure*}

\begin{figure*}
\includegraphics[width=90mm]{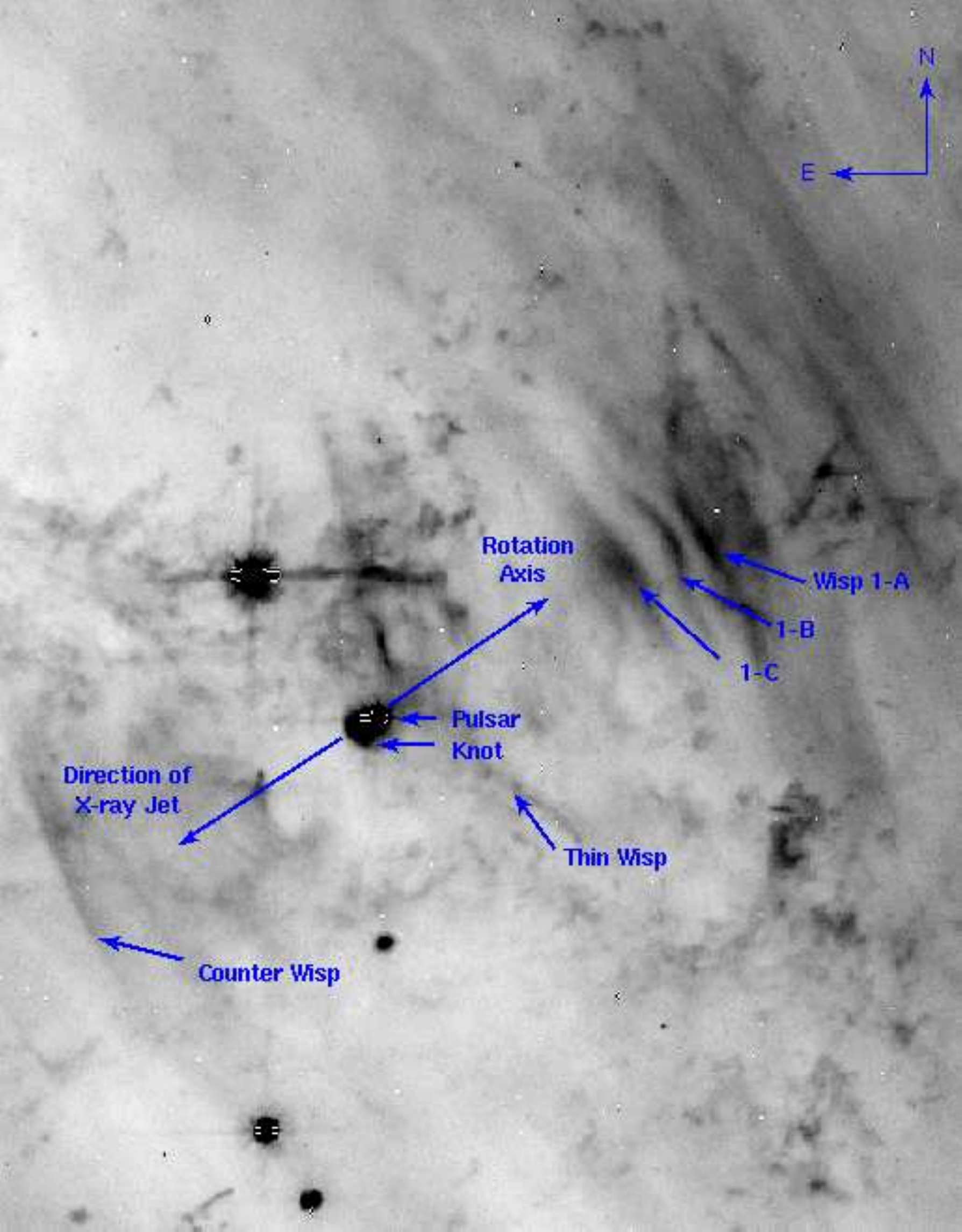}
\caption{HST/ACS image of the vicinity of the Crab pulsar (2005 Nov 25, FOV $\approx25\arcsec\times33\arcsec$), with features labelled for discussion.}
\label{figure4}
\end{figure*}

The raw Hubble Space Telescope/Advanced Camera for Surveys (HST/ACS) polarisation science frames of the Crab Nebula were obtained from the Mikulski Archive for Space Telescopes (MAST). The dataset consists of 13 observations of the nebula taken in three different polarisers (0\degr, 60\degr, and 120\degr) between 2003 August and 2005 December (Proposal ID: 9787) (see Table 1). The Wide Field Camera (WFC) detector, called ACS/WFC, employs a mosaic of two $4096\times2048$ Scientific Imaging Technologies (SITe) CCDs, with a pixel-scale of $\sim0.05$ arcsecond/pixel, covering a nominal FOV $\sim202\arcsec\times202\arcsec$ \citep{Pavlovsky}. For these observations, with the polarisers in place, the FOV was $\approx102\arcsec\times102\arcsec$. The filter used was F606W ($\lambda=590.70$ nm, $\Delta\lambda=250.00$ nm). The raw images, which had already been flat-fielded, were geometrically aligned, combined and averaged with cosmic-ray removal using IRAF (see Figure 1). We used a total of five field stars and the IRAF task \textit{ccmap} and the \textit{2MASS} catalogue to fit the astrometry. The pulsar was found at $\rm \alpha=05^{\rm h} 34 ^{\rm m} 31\fs930\pm0\fs001$, $\rm \delta=+22\degr00\arcmin51\farcs990\pm0\farcs110$, whilst the synchrotron knot, located 0\farcs65 SE of the pulsar \citep{Hester} is found at $\rm \alpha=05^{\rm h}34^{\rm m}31\fs980\pm0\fs001$, $\rm \delta=+22\degr00\arcmin51\farcs630\pm0\farcs110$ (the errors denote the rms of the astrometric fit). For each set of observations, the images taken in the different polarisers were analysed by the IMPOL\footnote{http://www.stecf.org/software/IRAFtools/stecf-iraf/impol} software \citep{Walsh}, which produces polarisation maps (see Figures 2 and 3).

In order to determine the polarimetry, aperture photometry was first performed on the pulsar and synchrotron knot in each image using the IRAF task \textit{phot}. The pulsar is saturated in each frame per epoch. \citet{Gilliland} describes the well behaved response of the ACS, and shows that electrons are conserved after saturation. The response of the ACS CCDs remains linear up to and beyond the point of saturation provided one uses a GAIN value that samples the full well depth.  For ACS this is a GAIN equal to 2 $\rm e^{-1}/ADU$, which is the GAIN setting used for these observations. Over a range of almost 4 magnitudes, photometry remains linear to $<1$\%. One can perform aperture photometry of isolated point sources by summing over all the pixels that were bled into \citep{Pavlovsky}. We tested this method by performing aperture photometry on the pulsar and Trimble 28. We used images taken at the same epoch as the polarimetric observations (2005 September to Decemeber) in the F550M filter ($\lambda=558.15$ nm, $\Delta\lambda=54.70$ nm) but with no polariser in place. We computed the visual magnitudes of both targets and found that the values are consistent with those of \citet{Sandberg}, once the different pass bands are taken into account.

We used an aperture of radius 0\farcs25 to measure the flux from the pulsar. The sky counts were measured using an annulus of width $\approx$ 0\farcs1, located 0\farcs15 beyond the central aperture. We added to this flux the flux from the pixels that were bled into. An aperture of radius 0\farcs15 was used to measure the flux from the synchrotron knot. The sky counts were measured in a region close to the pulsar and knot.

The Stokes vectors were then calculated using the following formulae:

\begin{equation}
I = \frac{2}{3} \ [r(0) + r(60) + r(120)]\\
\end{equation}

\begin{equation}
Q = \frac{2}{3} \ [2r(0) - r(60) - r(120)]\\
\end{equation}

\begin{equation}
\hspace{-0.85cm} U = \frac{2}{\sqrt{3}} \ [r(60) - r(120)] \\
\end{equation}
\\ \\
where r(0), r(60), and r(120) are the calibrated count rates in the 0, 60, and 120 degrees polarised images respectively \citep{Pavlovsky}.


\subsection{Computing the degree of linear polarisation of a target}

The degree of linear polarisation (P.D.) is calculated using the Stokes vectors, and factors which correct for cross-polarisation leakage in the polarising filters. This correction is useful for the POLUV filters; values for the the parallel and perpendicular transmission coefficients (T$_{\rm par}$ and T$_{\rm perp}$) can be found in Figure 5.4 of the ACS Instrument Handbook \citep{Pavlovsky}. These corrections together with the calibration of the source count rates removes the instrumental polarisation of the WFC ($\sim2\%$) (see Eqn. 4).


\subsection{Computing the polarisation position angle on the sky of the polarisation E-vector}

The position angle (P.A.) is calculated using the Stokes vectors, the roll angle of the HST spacecraft (PA$\_$V3 in the data header files), and $\chi$, which contains information about the camera geometry that is derived from the design specifications; for the WFC $\chi$= -38.2 degrees (see Eqn. 5). \\ \\

\begin{equation}
\rm P.D. = \frac{\sqrt{Q^{2} + U^{2}}}{I} \ \frac{T_{\rm par} + T_{\rm perp}}{{T_{\rm par} - T_{\rm perp}}} \times 100 \\ 
\end{equation}

\vspace{0.5cm}

\begin{equation}
\hspace{-0.4cm}\rm P.A. = \frac{1}{2} \ tan^{-1}\left(\frac{U}{Q} \right) + PA\_V3 + \chi \\ 
\end{equation} \\

An important property of polarisation that needs to be considered during analysis is that of bias. This is due to instrumental errors which tend to increase the observed polarisation of a target from its true polarisation. The effect is negligible when $\eta = \rm p \times S/N$ is high ($> 10$), where p is the fractional polarisation of the target, and S/N is the signal-to-noise per image. See for example Fig. 4 of \citet{Sparks}. Since the targets are in the high $\eta$ regime, the debiasing correction is small and therefore we omit it. We note that for stars 3 and 4, which have low $\eta$ values, that there will be a systematic over estimate of the polarisation, see \citet{Simmons} and \citet{Sparks}. However, as these have a polarisation consistent with zero no further analysis was performed on them to remove bias.

\citet{Naghizadeh-Khouei} investigated the statistical behaviour of the position angle of linear polarisation using both numerical integrations and data simulations. They found that the distribution of the angle is essentially Gaussian for $\eta > 6$. Hence, we used the formulae of \citet{Serkowski58,Serkowski62} for our error analysis. We propagated the errors in the count rates to obtain errors for the Stokes vectors I, Q, and U. Lastly, the errors in the Stokes vectors were propagated through the equations for the errors in the degree of polarisation (Eqn. 6) and position angle (Eqn. 7). As negative polarisation is impossible, we used asymmetric errorbars for stars 3 and 4. Below are the formulae used for calculating the errors in the degree of polarisation and position angle:

\begin{equation}
\rm \frac{\sigma_{P.D.}}{P.D.} = \sqrt{\frac{Q^{2}\sigma_Q^{2} + U^{2}\sigma_U^{2}}{Q^{2} + U^{2}} + \left(\frac{\sigma_I}{I} \right)^{2}}  \\ 
\end{equation}

\vspace{0.5cm}

\begin{equation}
\hspace{-2.0cm}\rm \sigma_{P.A.} = 28.65\degr \frac{\sigma_{P.D.}}{P.D.}\\ 
\end{equation}
\\
where $\sigma_{\rm I}$, $\sigma_{\rm Q}$, and $\sigma_{\rm U}$ are the errors in Stokes vectors I, Q, and U, respectively. These errors take into account those introduced by instrumention and systematics.

The polarisation of the synchrotron wisps was also studied (see Tables 2 and 3). To measure the total flux of each wisp, we summed the flux from a series of apertures (r $\approx$ 0\farcs3) placed along the extent of each wisp. We adopt the standard nomenclature as discussed by \cite{Scargle}, who noted their temporal variability and strong polarisation (see Figure 4). We have accounted for their temporal motion in our analysis. For the sky background subtraction we use the same region of the nebula as used for the knot. Since the contribution of zodiacal and scattered light to the background is low, we therefore ignore the effect of the background polarisation in our analysis.

As a guide to our analysis, a number of fore/background stars in the nebula (see Figure 3) were also analysed  to confirm the methodology which we used, and to cross-check for any systematics. Stars 3 and 4 are not saturated in each frame per epoch, but stars 1, 2, and Trimble 28 are saturated. Therefore we employed the same photometric method as used for the pulsar. We used an aperture of radius 0\farcs35 to measure the flux from each star. The sky counts were measured using an annulus of width $\approx$ 0\farcs1, located 0\farcs15 beyond the central aperture. We have found that all of the stars are consistent, within the errors, with unpolarised sources.

We have omitted the results of the analysis of the 2003 August dataset. The sky background is higly variable in one of the raw images. This then causes errors when one calculates the polarisation of the wisps and plots the polarisation maps for this epoch. We also note that for the 2005 December dataset that the roll-angle of the spacecraft was significantly different to other observations. For this dataset the diffraction spike from the pulsar crosses the knot. Furthermore, we note that for the December 14th dataset the full Moon was 9 degrees away from the pulsar. This might have introduced spurious background levels which would have impacted upon the polarisation of the faint extended sources such as the wisps.

We have investigated the effects of photometric losses due to charge transfer efficiency (CTE) in the CCDs of the WFC. The effect reduces the apparent brightness of sources, and it requires a photometric correction to restore the measured integrated counts to their \lq\lq true\rq\rq\ value. The ACS team claim that there is no evidence of photometric losses due to CTE for WFC data taken after 2004. Nonetheless, we applied the correction for CTE (see Eqn. 8) to our photometry and found that it does not change the results of the polarimetry. Below is the formula for the correction for CTE loss. This value is then added to the measured flux.\\ 

\begin{equation}
\hspace{-0.2cm} YCTE = 10^{A} \times Sky^{B} \times Flux^{C} \times \frac{Y}{2048} \times \frac{(MJD-52333)}{365} \\ 
\end{equation} 
\\ 
where MJD is the modified julian date of the observation, and reflects the linear degradation of the CCD with time. The parameters A, B, and C are found in Table 6.1 of the ACS Instrument Handbook \citep{Pavlovsky}.
 
In order to determine the performance of the ACS as a polarimeter, the ACS team have modeled the complete instrumental effects and the calibration together. This is done so as to quantify the impacts of the remaining uncalibrated systematic errors. They claim that the fractional polarisations  will be uncertain at the one-part-in-ten level (e.g. a 20\% polarisation has an uncertainty of 2\%) for strongly polarised targets; and at about the 1\% level for weakly polarised targets. The position angles will have an uncertainty of about 3\degr. This is in addition to uncertainties which arise from photon statistics \citep{Pavlovsky}. They then checked this calibration against polarised standard stars (~5\% polarised) and found it to be reliable within the quoted errors \citep{Cracraft}.


\subsection{Photometry and morphology of the knot
   in unpolarised light}

We also retrieved from the MAST archive a series of 12 ACS/WFC datasets, collected through the F550M filter ($\lambda=558.15$ nm, $\Delta\lambda=54.70$ nm) at the same epoch as the polarimetric observations (from 2005, September 6 to 2005, December 14). Each observation consists of a sequence of two images collected in a single orbit, to allow for cosmic-ray rejection. We retrieved pipeline-calibrated, {\em drizzled}\footnote{single, calibrated ACS images were combined using the {\tt multidrizzle} software, which also produces a mosaic image of the two ACS chips and applies a correction for the geometric distortions of the camera.} images from the archive. Total exposure times range from  1950 s to 2300 s per epoch (see Table 1).
We superimposed the images on the first-epoch one by using the coordinates of 30 non saturated field sources as a reference grid. The rms accuracy was better than 0.07 pixels per coordinate. We performed multi-epoch photometry of the knot with the {\tt Sextractor} software \citep{Bertin}. We used an implementation of the Kron method \citep{Kron}, which measures the flux of an object within an optimised elliptical aperture, evaluated using the second moments of the object's brightness distribution. The parameters of the Kron ellipse (center, semiaxes, and orientation) also yield a measure of the object coordinates and morphology, which is useful for the case of the knot, which is possibly variable in both position and shape as a function of time. The measured count rate was  converted to flux using the standard ACS photometric calibration tabulated in the image headers. Correction for CTE losses proved to have a negligible effect. 

Since the knot is a diffuse source located very close to a much brighter and saturated point source (the Crab pulsar), particular care was devoted to estimate systematic errors possibly affecting the flux measurements. To this aim, we have performed simulations with the ESO/Midas software\footnote{http://www.eso.org/sci/software/esomidas/}, adding to the ACS images a \lq\lq synthetic knot\rq\rq. A 2-dimensional Gaussian function was used to generate the artificial source, setting one of the symmetry axes aligned to the pulsar spin-axis. The synthetic knot was positioned to the NW of the pulsar, along the direction of the pulsar spin-axis, at an angular distance comparable to that of the true knot. By varying the flux and the position of the artificial source we evaluated the uncertainty on the flux measurement to be $\sim2.5\%$. 

We also measured the fluxes of a sample of non-saturated stars in the field as a further assessment of the stability of photometry in the variable background of the Crab nebula (see Figure 3).


\section{Results}

Included here are the measurements of the degree of polarisation and position angle of each target per epoch (see Tables 2--5 inclusive). We have plotted the degree of polarisation and position angles for each target as a function of time (see Figures 5 and 6). Using a $\chi^{2}$ goodness-of-fit, we found no significant variation (at the 95\% confidence level) in the polarisation of the pulsar, knot, and wisps over the 3 month period of these observations. As a final comparison, we present the mean values. These are the values obtained from using the weighted mean and error of the degree of polarisation and position angle (see Table 6). Stars 3 and 4 have asymmetric error bars. Hence, we use the method of \citet{Barlow} for calculating the weighted mean for asymmetric error bars. As seen from Figure 5 and 6 and Table 6, the 2005 December dataset shows evidence of a possible variation of the knot polarisation at the $2 \sigma$ level. This variation is due neither to a known systematic effect nor to the contribution of the diffraction spikes from the pulsar (see Sectn. 2.2), which only affect the knot's flux by $\la 2\%$. Future polarimetry observations on a longer time span will help us to address the possible knot variability. Similarly, we note that the polarisation of Wisps 1-A and 1-B also shows a possible variation at the $2 \sigma$ level in the 2005 December dataset. As discussed in Sectn. 2.2, this may be partially ascribed to the enhanced Moon contribution to the background. This possible variation may be also ascribed to the unresolved contribution of the bright torus in the nebula, to which Wisps 1-A and B have moved closest in the 2005 December observations.

The polarisation maps (Figure 2 and 7) show the variation of the polarisation throughout the inner nebula and particularly in the vicinity of the pulsar itself. Each vector has magnitude equal to the degree of polarisation, and its orientation is the position angle at that point. Such maps allow one to visualise the direction of the magnetic field lines within the nebula. One can distinctly see the overall structure of the inner nebula, the degree of polarisation of the knots and the synchrotron emission. In particular, the filaments are unpolarised and the structures that are visible in polarised light (Figure 3) do not map exactly the continuum (Figure 1).

The F550M band images were used to study with high accuracy possible displacements of the knot. The light curve of the knot is shown in Figure 8, compared to the one of the reference stars. The knot is seen to brighten by $\sim40\%$ on a 60-day time scale, then to fade to its initial flux level. Reference stars, conversely, do not display any significant variability. Focusing on the knot, we note that changes in flux are accompanied by shifts in position, the knot centroid being closer to the pulsar when the feature is brighter (the displacement between maximum and minimum flux is $0\farcs075\pm0.025$). There is no statistically significant evidence of a change in the FWHM of the knot as a function of time. We checked our photometric results by repeating the analysis with simple aperture photometry, using an aperture of 0\farcs15 positioned on the knot centroid (as measured in each epoch). Such an exercise yielded consistent results ($\sim40\%$ brightening in two months), confirming the significant variability in flux.


 \begin{table*}
 \caption{Degree of linear polarisation (\%) of the Crab pulsar, synchrotron knot, and wisps as a function of time. Wisp 1-A is unresolved from 2005 Sep 06 -- 2005 Oct 12 inclusive.}
 \label{symbols}
 \begin{tabular}{lccccccc}
  \hline
  Date & Crab & Knot & Wisp 1-A & Wisp 1-B & Wisp 1-C & Thin Wisp & Counter Wisp\\
  \hline
  2005 Sep 06   &4.9$\pm$1.0   &59.4$\pm$7.3   &			&41.7$\pm$3.8   &32.1$\pm$4.5   &36.1$\pm$4.0  &31.0$\pm$4.3\\
  2005 Sep 15   &5.0$\pm$1.0   &60.0$\pm$7.0   &  			&45.0$\pm$4.0   &34.5$\pm$4.6   &37.5$\pm$4.4  &35.6$\pm$4.6\\
  2005 Sep 25   &5.4$\pm$1.0   &62.1$\pm$7.1   &			&44.2$\pm$4.3   &40.7$\pm$4.7   &38.5$\pm$4.7  &40.1$\pm$5.1\\
  2005 Oct 02   &4.9$\pm$1.0   &62.4$\pm$6.9   &			&41.3$\pm$4.5   &39.7$\pm$4.7   &37.2$\pm$4.5  &39.2$\pm$5.1\\
  2005 Oct 12   &4.6$\pm$1.0   &60.3$\pm$6.6   &			&43.4$\pm$4.5   &36.8$\pm$4.8   &32.6$\pm$4.7  &37.5$\pm$5.2\\
  2005 Oct 22   &4.3$\pm$1.0   &61.1$\pm$6.6   &43.6$\pm$4.7		&46.2$\pm$4.5   &37.5$\pm$5.0   &36.9$\pm$4.7  &45.1$\pm$5.4\\
  2005 Oct 30   &4.9$\pm$1.0   &61.4$\pm$6.4   &44.1$\pm$4.6		&44.9$\pm$4.6   &35.2$\pm$4.8   &34.8$\pm$4.9  &42.7$\pm$5.6\\
  2005 Nov 08   &5.1$\pm$1.0   &60.9$\pm$6.2   &41.6$\pm$4.3		&45.6$\pm$4.5   &39.3$\pm$4.6   &35.0$\pm$5.0  &41.0$\pm$5.6\\
  2005 Nov 16   &5.6$\pm$1.0   &59.9$\pm$6.4   &41.5$\pm$4.4		&46.0$\pm$4.8   &42.4$\pm$4.7   &38.1$\pm$5.3  &47.9$\pm$6.3\\
  2005 Nov 25   &5.8$\pm$1.0   &59.9$\pm$6.9   &41.4$\pm$4.4		&40.1$\pm$4.8   &38.9$\pm$4.5   &38.6$\pm$5.5  &41.7$\pm$5.9\\
  2005 Dec 05   &5.9$\pm$1.0   &42.8$\pm$6.2   &32.1$\pm$3.7		&40.9$\pm$4.3   &38.9$\pm$4.5   &38.1$\pm$4.7  &43.2$\pm$5.1\\
  2005 Dec 14   &5.1$\pm$1.0   &43.9$\pm$6.6   &21.2$\pm$3.7		&25.2$\pm$3.7   &40.5$\pm$4.5   &35.4$\pm$5.1  &39.1$\pm$5.5\\
  \hline
 \end{tabular}
\end{table*}

\begin{table*}
 \caption{Polarisation position angles ($^{\circ}$) of the Crab pulsar, synchrotron knot, and wisps as a function of time. Wisp 1-A is unresolved from 2005 Sep 06 -- 2005 Oct 12 inclusive.}
 \label{symbols}
 \begin{tabular}{lcccccccccc}
  \hline
  Date & Crab & Knot & Wisp 1-A & Wisp 1-B & Wisp 1-C & Thin Wisp & Counter Wisp\\
  \hline
  2005 Sep 06   &103.5$\pm$5.9   &123.9$\pm$3.5   &			&124.8$\pm$2.6  &129.1$\pm$4.0  &128.0$\pm$3.1  &131.5$\pm$3.9\\
  2005 Sep 15   &103.2$\pm$5.9   &124.7$\pm$3.4   &			&123.7$\pm$2.5  &124.7$\pm$3.8  &125.4$\pm$3.3  &128.1$\pm$3.7\\
  2005 Sep 25   &106.6$\pm$5.4   &125.0$\pm$3.3   &			&125.7$\pm$2.8  &125.6$\pm$3.3  &125.7$\pm$3.5  &128.6$\pm$3.7\\
  2005 Oct 02   &103.5$\pm$5.8   &125.1$\pm$3.2   &			&126.5$\pm$3.1  &125.6$\pm$3.4  &126.3$\pm$3.5  &128.6$\pm$3.7\\
  2005 Oct 12   &102.6$\pm$6.4   &124.9$\pm$3.2   &			&129.5$\pm$3.0  &129.8$\pm$3.7  &127.7$\pm$4.2  &131.0$\pm$3.9\\
  2005 Oct 22   &109.7$\pm$6.7   &125.7$\pm$3.1   &124.8$\pm$3.1 	&127.2$\pm$2.8  &125.5$\pm$3.8  &124.2$\pm$3.6  &126.7$\pm$3.5\\
  2005 Oct 30   &104.6$\pm$5.8   &125.7$\pm$3.0   &125.0$\pm$3.0 	&130.3$\pm$3.0  &127.8$\pm$3.9  &126.8$\pm$4.0  &129.6$\pm$3.8\\
  2005 Nov 08   &106.6$\pm$5.7   &125.4$\pm$2.9   &123.4$\pm$3.0 	&130.6$\pm$2.8  &129.3$\pm$3.3  &127.1$\pm$4.1  &130.7$\pm$3.9\\
  2005 Nov 16   &108.0$\pm$5.1   &124.7$\pm$3.1   &124.5$\pm$3.0 	&130.0$\pm$3.0  &129.6$\pm$3.2  &128.0$\pm$4.0  &131.5$\pm$3.8\\
  2005 Nov 25   &102.2$\pm$5.1   &125.4$\pm$3.3   &125.5$\pm$3.0  	&127.4$\pm$3.5  &130.2$\pm$3.3  &125.9$\pm$4.1  &129.5$\pm$4.1\\
  2005 Dec 05   &107.5$\pm$4.9   &121.6$\pm$4.2   &122.0$\pm$3.3	&125.7$\pm$3.0  &130.5$\pm$3.2  &124.4$\pm$3.5  &127.3$\pm$4.4\\
  2005 Dec 14   &102.5$\pm$5.7   &119.9$\pm$4.3   &132.6$\pm$5.1	&131.4$\pm$4.2  &134.4$\pm$3.1  &139.2$\pm$4.2  &143.2$\pm$4.0\\
  \hline
 \end{tabular}
\end{table*}

\begin{table*}
 \caption{Degree of linear polarisation (\%) of Trimble 28 and the background stars a function of time.}
 \label{symbols}
 \begin{tabular}{lccccc}
  \hline
  Date & Trimble 28 & Star 1 & Star 2 & Star 3 & Star 4\\
  \hline
  2005 Sep 06   &1.0$\pm$0.7   &1.5$\pm$1.3		&0.9$\pm$0.9  		&1.6$^{+2.8}_{-1.6}$	&3.4$^{+3.7}_{-3.4}$\\
  2005 Sep 15   &1.1$\pm$0.6   &2.4$\pm$1.3		&1.2$\pm$0.9  		&2.6$^{+2.8}_{-2.6}$ 	&3.4$^{+3.7}_{-3.4}$\\
  2005 Sep 25   &1.7$\pm$0.7   &2.4$\pm$1.4		&1.0$\pm$0.9  		&2.7$^{+2.8}_{-2.7}$  	&3.3$^{+3.7}_{-3.3}$\\
  2005 Oct 02   &1.1$\pm$0.6   &3.1$\pm$1.3		&1.2$\pm$0.9  		&1.7$^{+2.8}_{-1.7}$  	&2.8$^{+3.6}_{-2.8}$\\
  2005 Oct 12   &0.9$\pm$0.7   &2.0$\pm$1.3		&0.9$\pm$0.9  		&2.0$^{+2.8}_{-2.0}$  	&2.2$^{+3.7}_{-2.2}$\\
  2005 Oct 22   &2.1$\pm$0.7   &2.5$\pm$1.3		&1.5$\pm$0.9  		&2.9$\pm$2.8  		&4.3$\pm$3.7\\
  2005 Oct 30   &0.7$\pm$0.6   &1.5$\pm$1.3		&1.4$\pm$0.9  		&2.0$^{+2.8}_{-2.0}$  	&2.7$^{+3.7}_{-2.7}$\\
  2005 Nov 08   &1.2$\pm$0.7   &2.3$\pm$1.4		&0.7$^{+0.9}_{-0.7}$  	&3.3$\pm$2.8  		&3.7$\pm$3.7\\
  2005 Nov 16   &1.6$\pm$0.7   &1.2$^{+1.3}_{-1.2}$	&1.6$\pm$0.9  		&2.1$^{+2.8}_{-2.1}$  	&1.8$^{+3.7}_{-1.8}$\\
  2005 Nov 25   &1.5$\pm$0.7   &2.0$\pm$1.4   		&1.5$\pm$0.9  		&2.5$^{+2.9}_{-2.5}$  	&3.5$^{+3.8}_{-3.5}$\\
  2005 Dec 05   &1.8$\pm$0.7   &2.2$\pm$1.4   		&1.2$\pm$0.9  		&2.2$^{+2.9}_{-2.2}$ 	&3.0$^{+3.8}_{-3.0}$\\
  2005 Dec 14   &1.9$\pm$0.7   &2.5$\pm$1.4   		&1.5$\pm$0.9  		&1.3$^{+2.9}_{-1.3}$  	&1.5$^{+3.7}_{-1.5}$\\
 \hline
 \end{tabular}
\end{table*}
 
\begin{table*}
 \caption{Polarisation position angles ($^{\circ}$) of Trimble 28 and the background stars as a function of time.}
 \label{symbols}
 \begin{tabular}{lccccc}
  \hline
  Date & Trimble 28 & Star 1 & Star 2 & Star 3 & Star 4\\
  \hline
  2005 Sep 06   &116.3$\pm$18.5	&149.5$\pm$25.8     &177.0$\pm$28.1    &139.3$\pm$49.5	&146.6$\pm$30.7\\
  2005 Sep 15   &170.8$\pm$17.6	&145.3$\pm$15.9     &174.4$\pm$22.2	&138.7$\pm$31.4	&136.2$\pm$30.7\\
  2005 Sep 25   &159.4$\pm$11.1	&148.2$\pm$12.6     &124.2$\pm$25.2	&145.7$\pm$30.5	&144.4$\pm$32.0\\
  2005 Oct 02   &178.5$\pm$16.3	&151.1$\pm$19.4     &1.0$\pm$20.5	&125.9$\pm$47.0	&148.1$\pm$36.9\\
  2005 Oct 12   &74.8$\pm$20.5		&151.1$\pm$18.6     &97.6$\pm$27.1	&148.9$\pm$39.7	&145.8$\pm$47.1\\
  2005 Oct 22   &152.4$\pm$8.0		&147.4$\pm$15.4     &138.8$\pm$17.1	&148.8$\pm$28.3	&148.6$\pm$24.5\\
  2005 Oct 30   &140.0$\pm$24.9	&153.8$\pm$26.3     &136.7$\pm$17.9	&150.8$\pm$41.0	&142.2$\pm$38.3\\
  2005 Nov 08   &155.6$\pm$15.1	&151.3$\pm$16.5     &124.2$\pm$36.8	&149.0$\pm$24.5	&139.5$\pm$29.0\\
  2005 Nov 16   &156.8$\pm$12.0	&144.5$\pm$33.2     &116.3$\pm$15.7	&140.2$\pm$38.7	&144.4$\pm$60.1\\
  2005 Nov 25   &142.6$\pm$12.2	&153.8$\pm$19.9     &136.6$\pm$17.0	&135.3$\pm$33.0	&149.6$\pm$30.9\\
  2005 Dec 05   &135.5$\pm$10.4   	&132.4$\pm$17.3     &144.2$\pm$22.2 	&146.4$\pm$37.4   	&146.1$\pm$36.2\\
  2005 Dec 14   &144.1$\pm$10.1  	&148.7$\pm$15.8     &164.9$\pm$17.0 	&150.4$\pm$64.5	&149.1$\pm$69.2\\
  \hline
 \end{tabular}
\end{table*}


\begin{center}
\begin{table*}
 \caption{Overall results for the degree of linear polarisation (\%) and position angle ($^{\circ}$). These are the weighted mean and error of the degree of polarisation and position angle.}
 \label{symbols}
 \begin{tabular}{@{}lcc}
  \hline
   & Polarisation Degree (\%) & Position Angle ($^{\circ}$) \\
  \hline
  \bf{Pulsar}      &5.2$\pm$0.3   &105.1$\pm$1.6\\
  Synchrotron Knot &59.0$\pm$1.9  &124.7$\pm$1.0\\
  Wisp 1-A         &39.8$\pm$1.6  &124.7$\pm$1.2\\
  Wisp 1-B         &43.0$\pm$1.3  &127.4$\pm$0.9\\
  Wisp 1-C         &38.5$\pm$1.3  &128.8$\pm$1.0\\
  Thin Wisp        &36.7$\pm$1.4  &127.1$\pm$1.0\\
  Counter Wisp     &40.6$\pm$1.5  &130.3$\pm$1.1\\
  Trimble 28       &1.6$\pm$0.2   &147.5$\pm$3.7\\
  Star 1  	   &2.3$\pm$0.4   &147.7$\pm$5.2\\
  Star 2   	   &1.2$\pm$0.3   &128.3$\pm$5.9\\
  Star 3  	   &1.6$\pm$0.7   &144.0$\pm$10.2\\
  Star 4  	   &2.2$\pm$1.0   &144.8$\pm$9.9\\
  \hline
 \end{tabular}
\end{table*}
\end{center}


\begin{figure*}
\centering
\subfloat{\includegraphics[width=45mm]{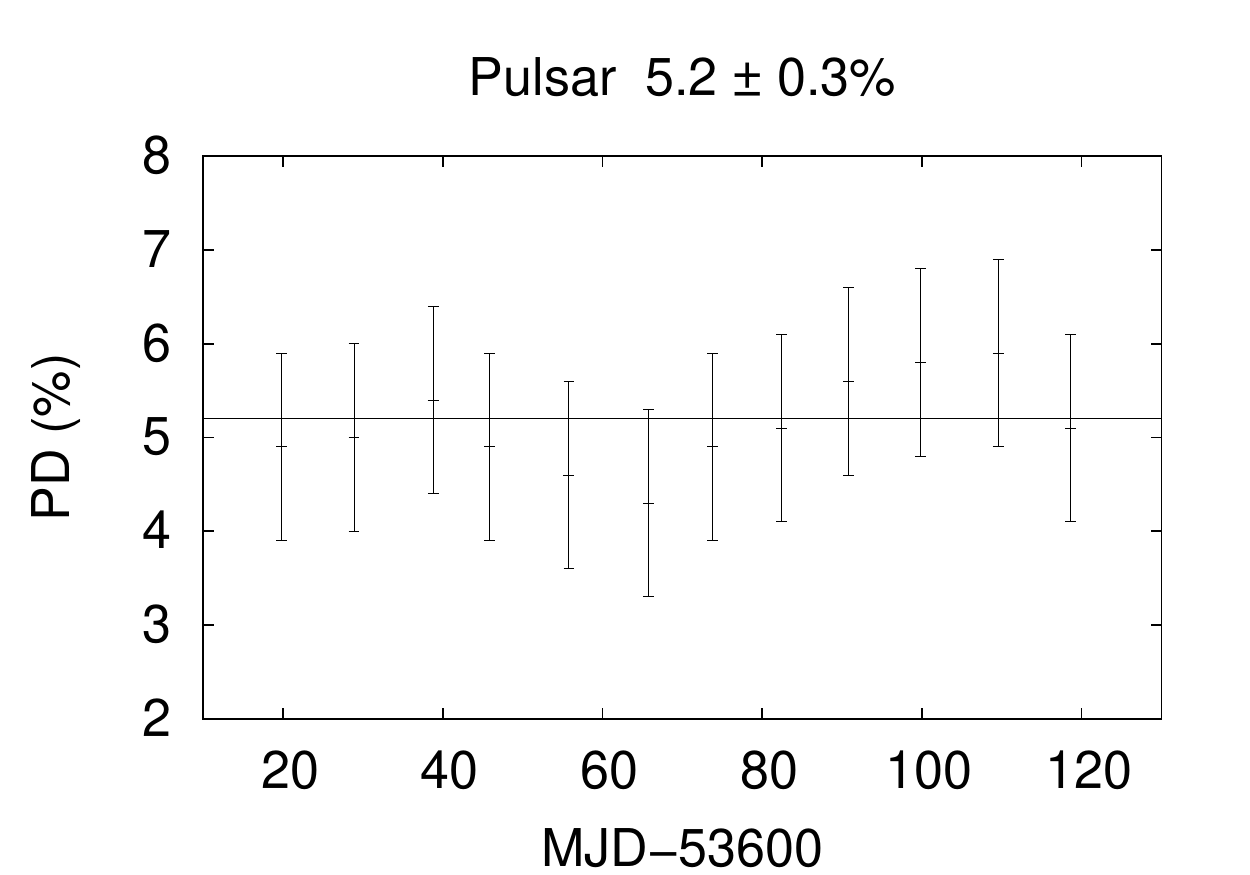}}
\subfloat{\includegraphics[width=45mm]{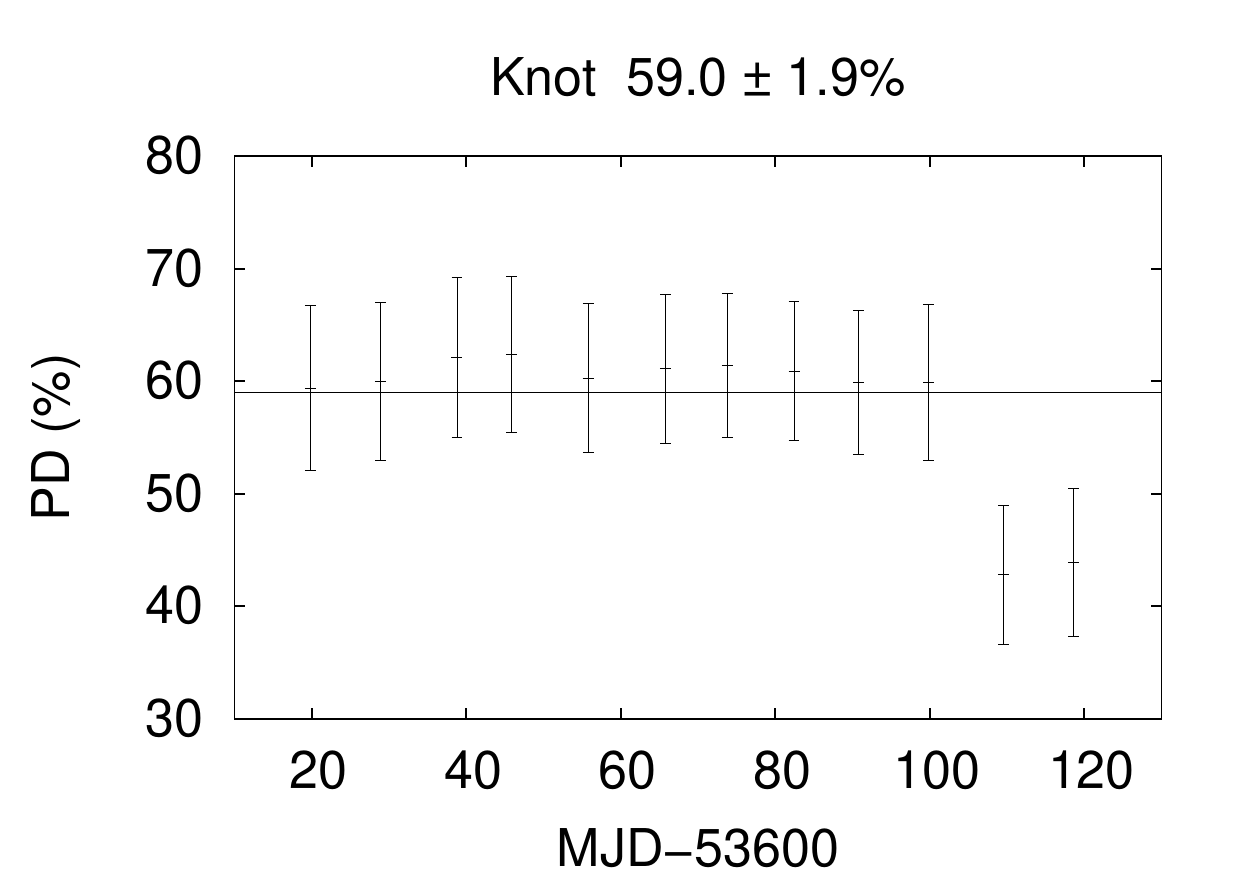}}
\subfloat{\includegraphics[width=45mm]{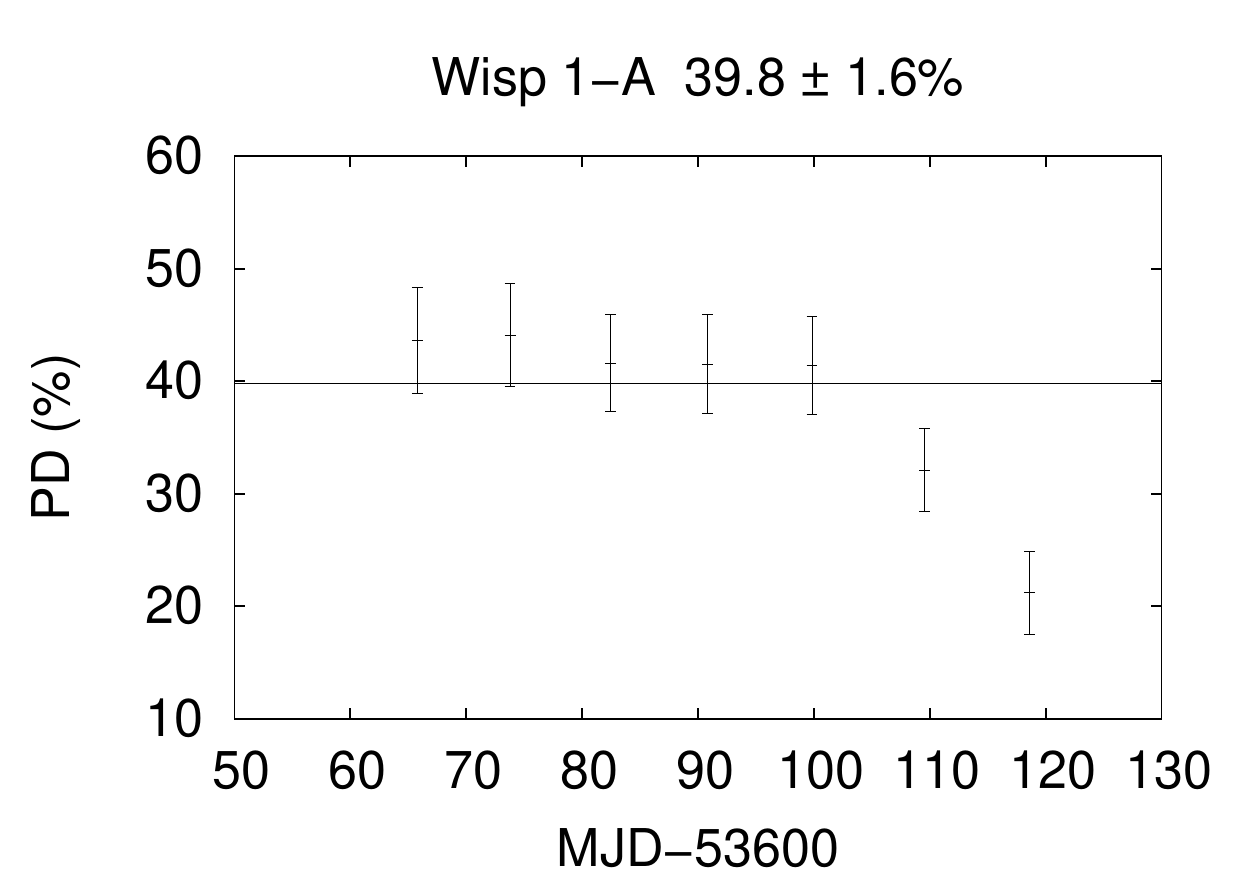}}
\subfloat{\includegraphics[width=45mm]{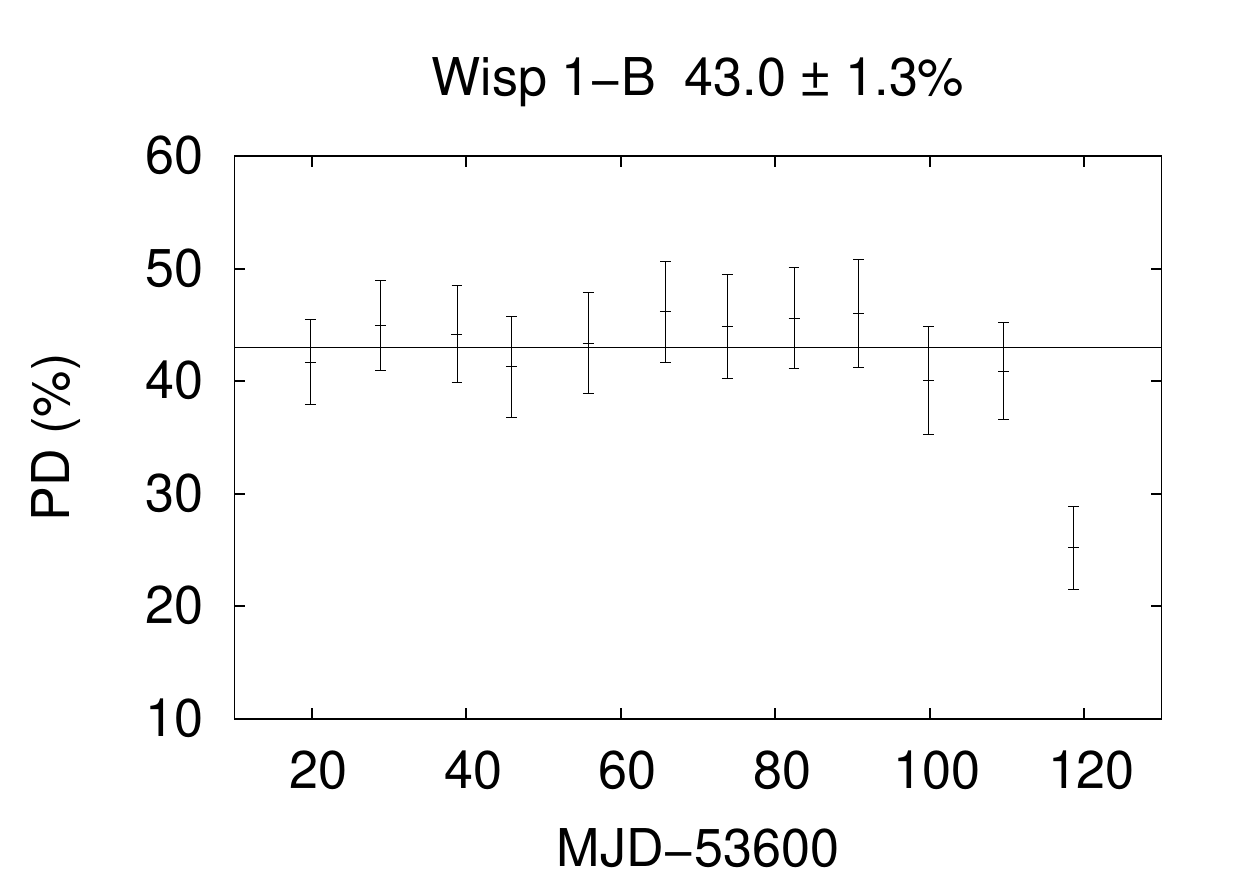}}

\subfloat{\includegraphics[width=45mm]{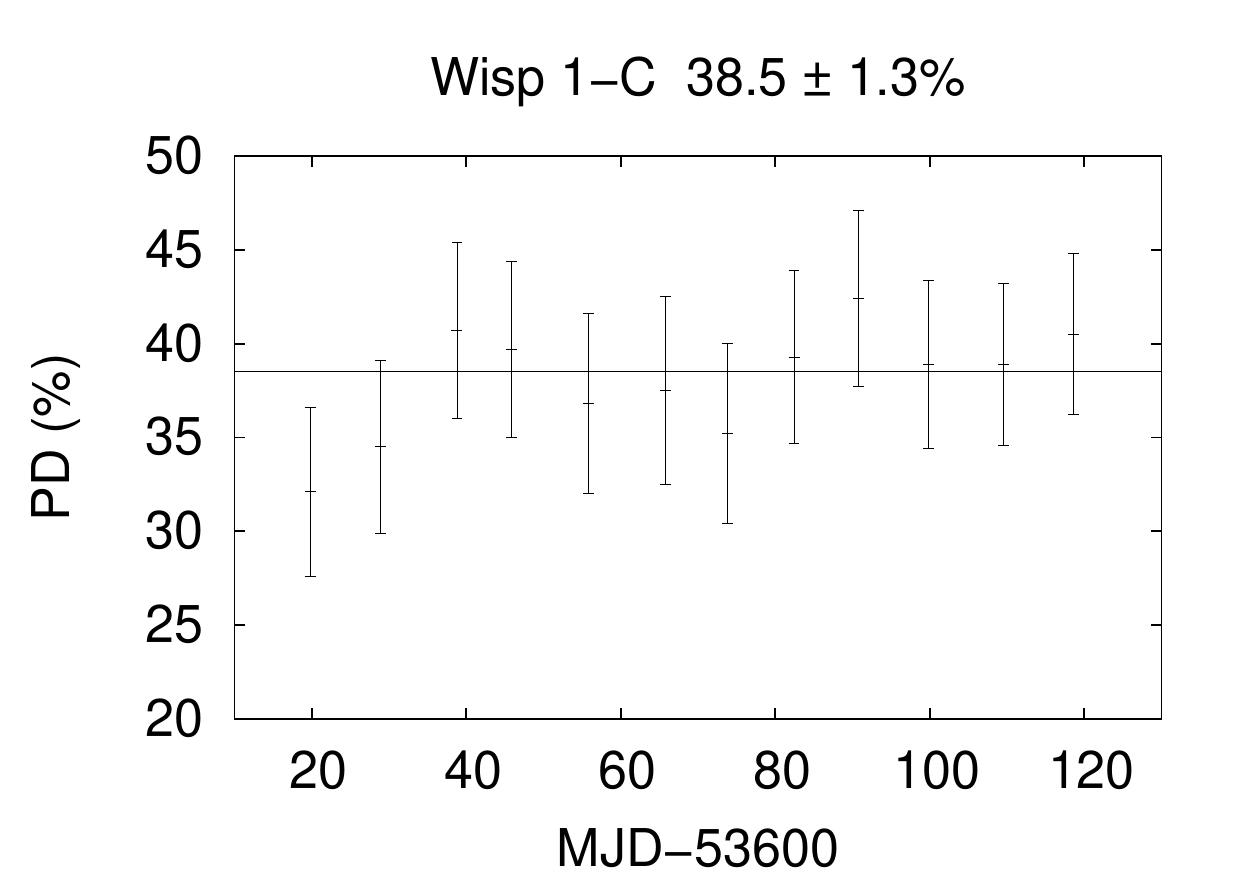}}
\subfloat{\includegraphics[width=45mm]{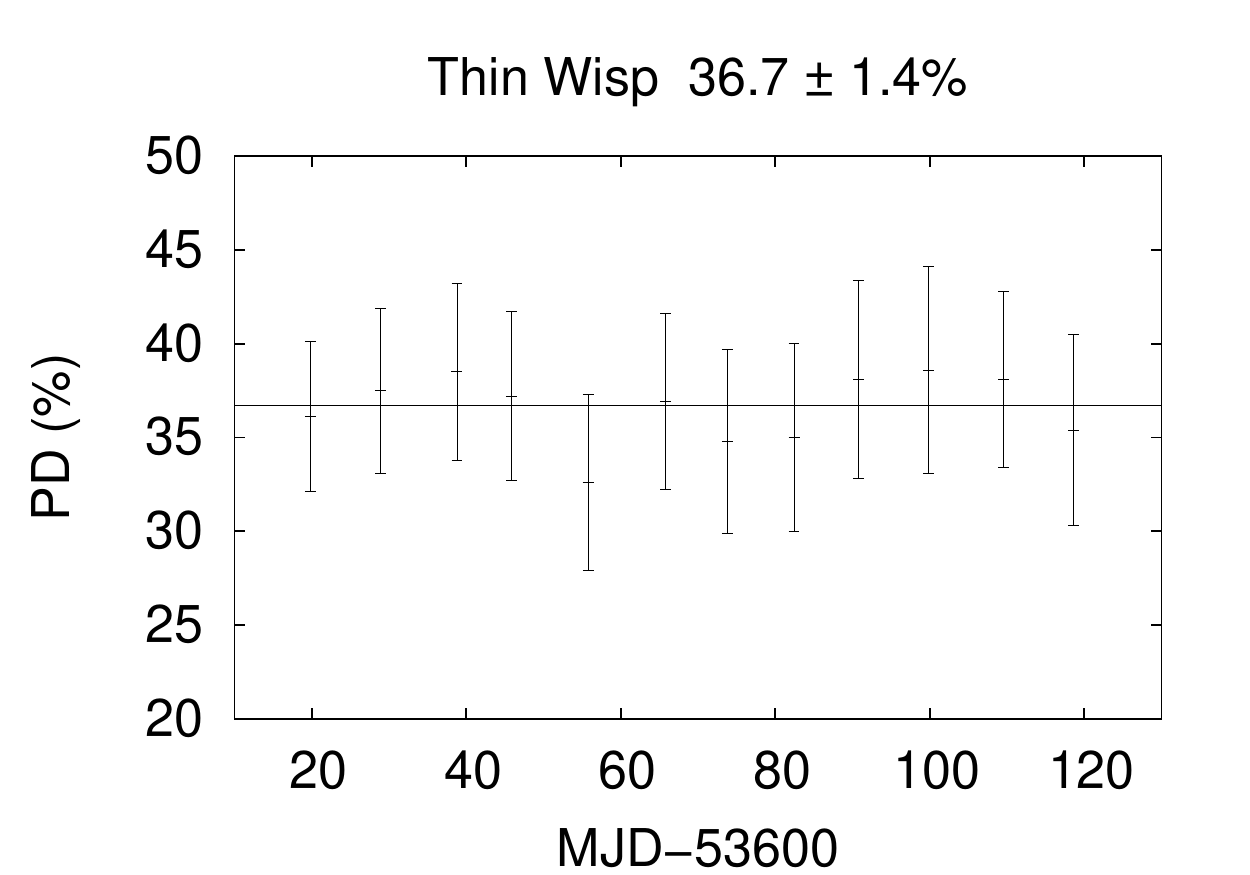}}
\subfloat{\includegraphics[width=45mm]{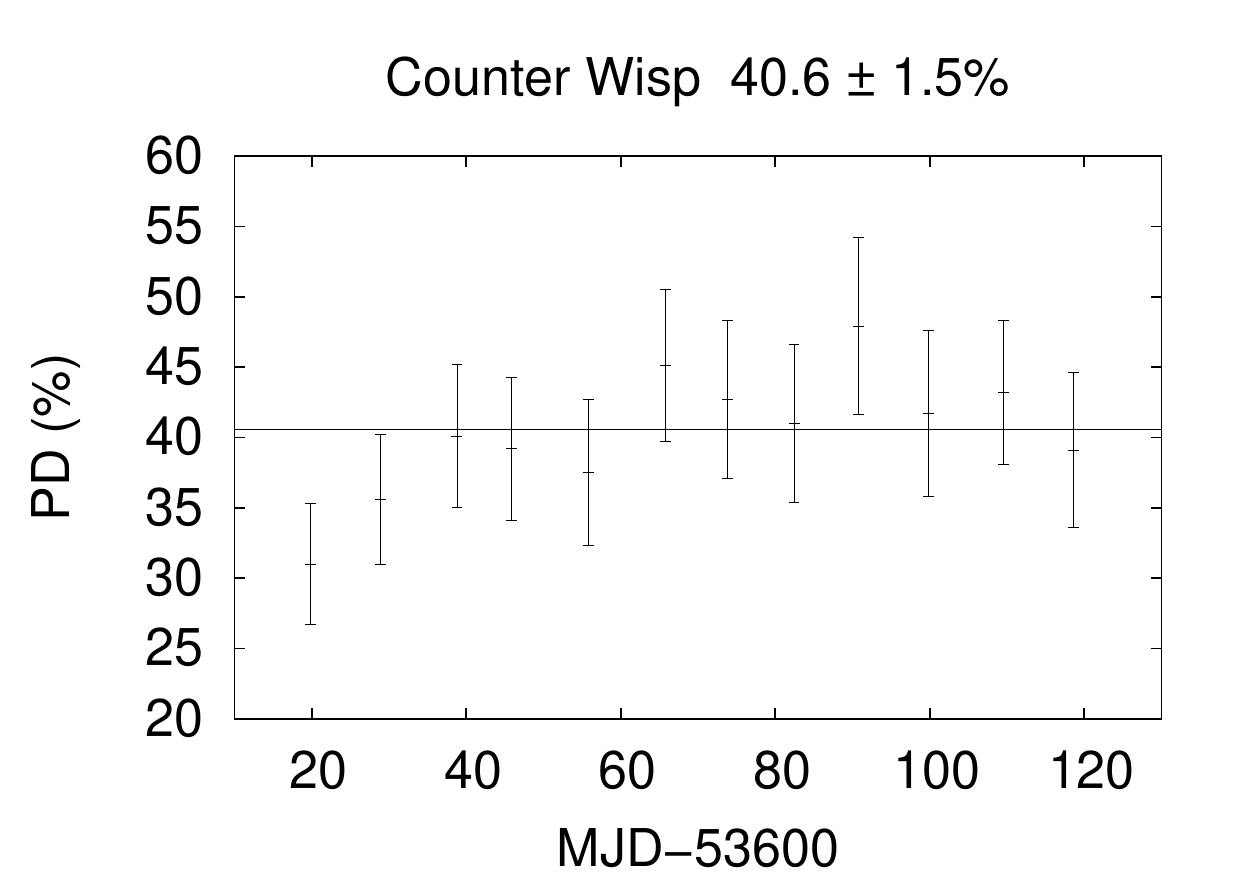}}
\subfloat{\includegraphics[width=45mm]{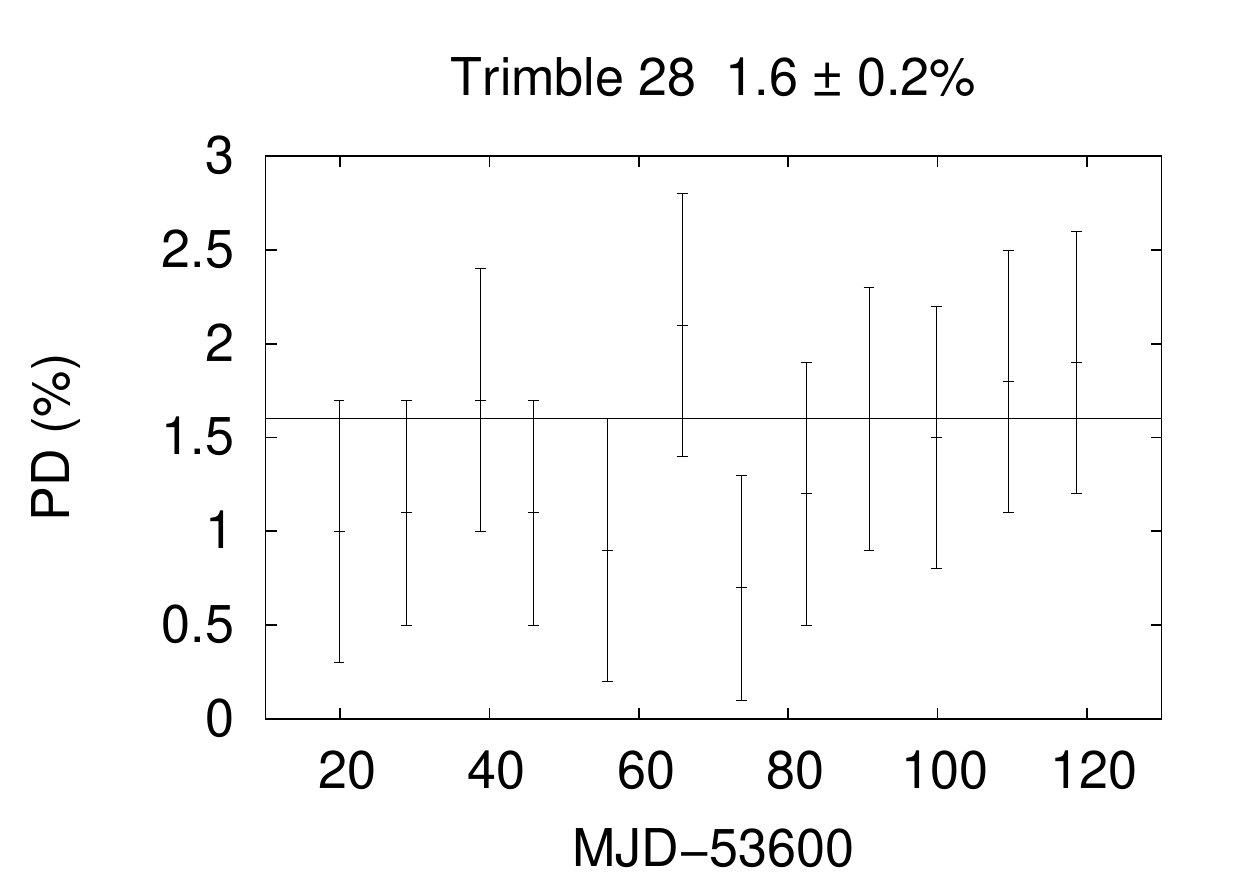}}

\subfloat{\includegraphics[width=45mm]{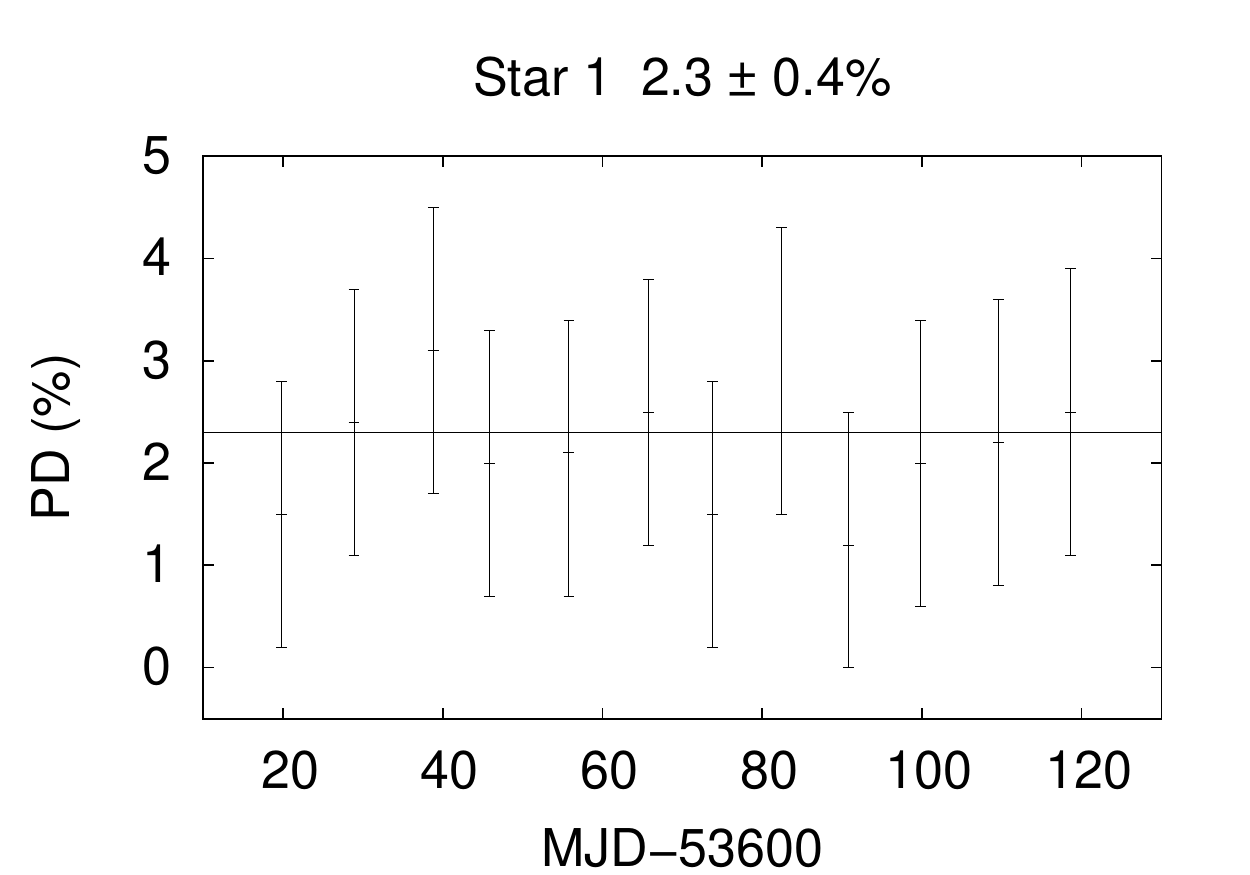}}
\subfloat{\includegraphics[width=45mm]{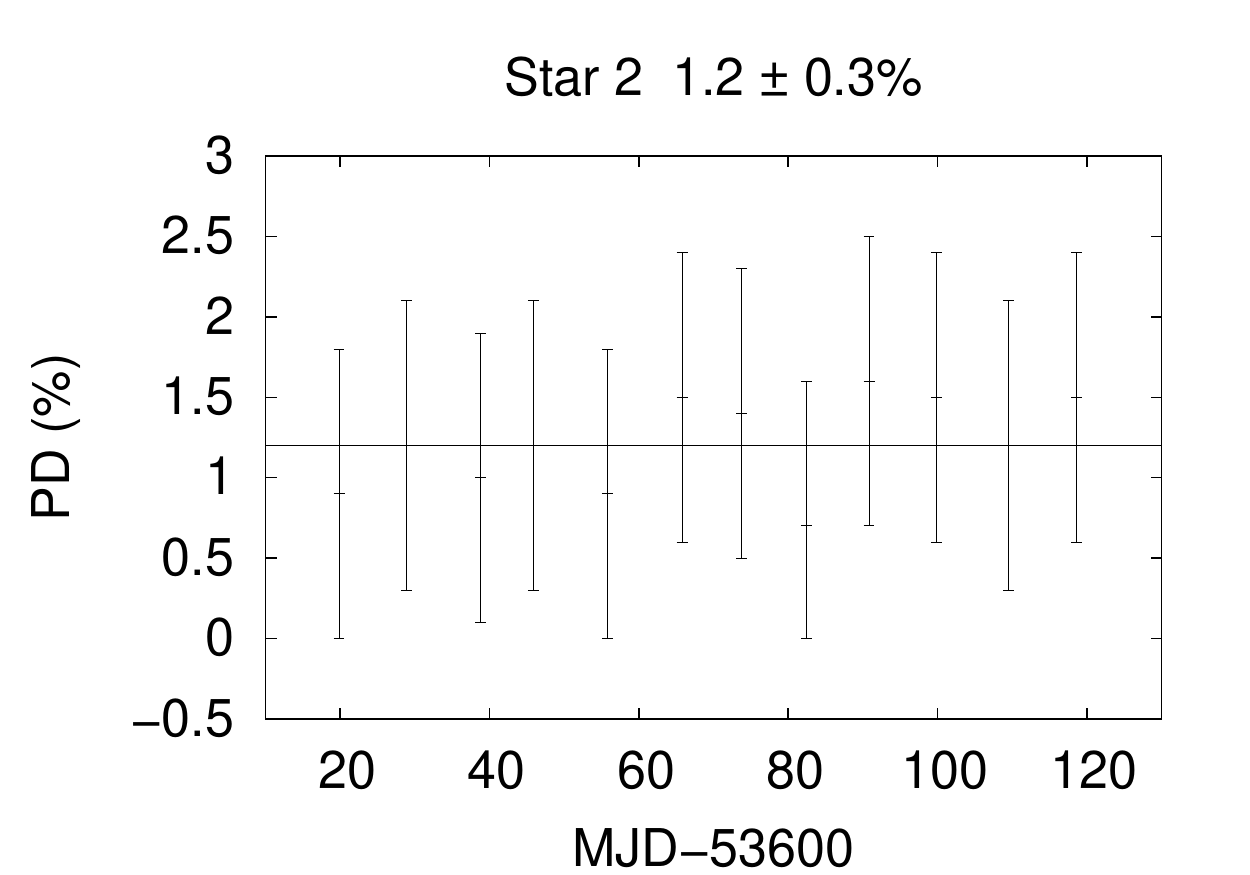}}
\subfloat{\includegraphics[width=45mm]{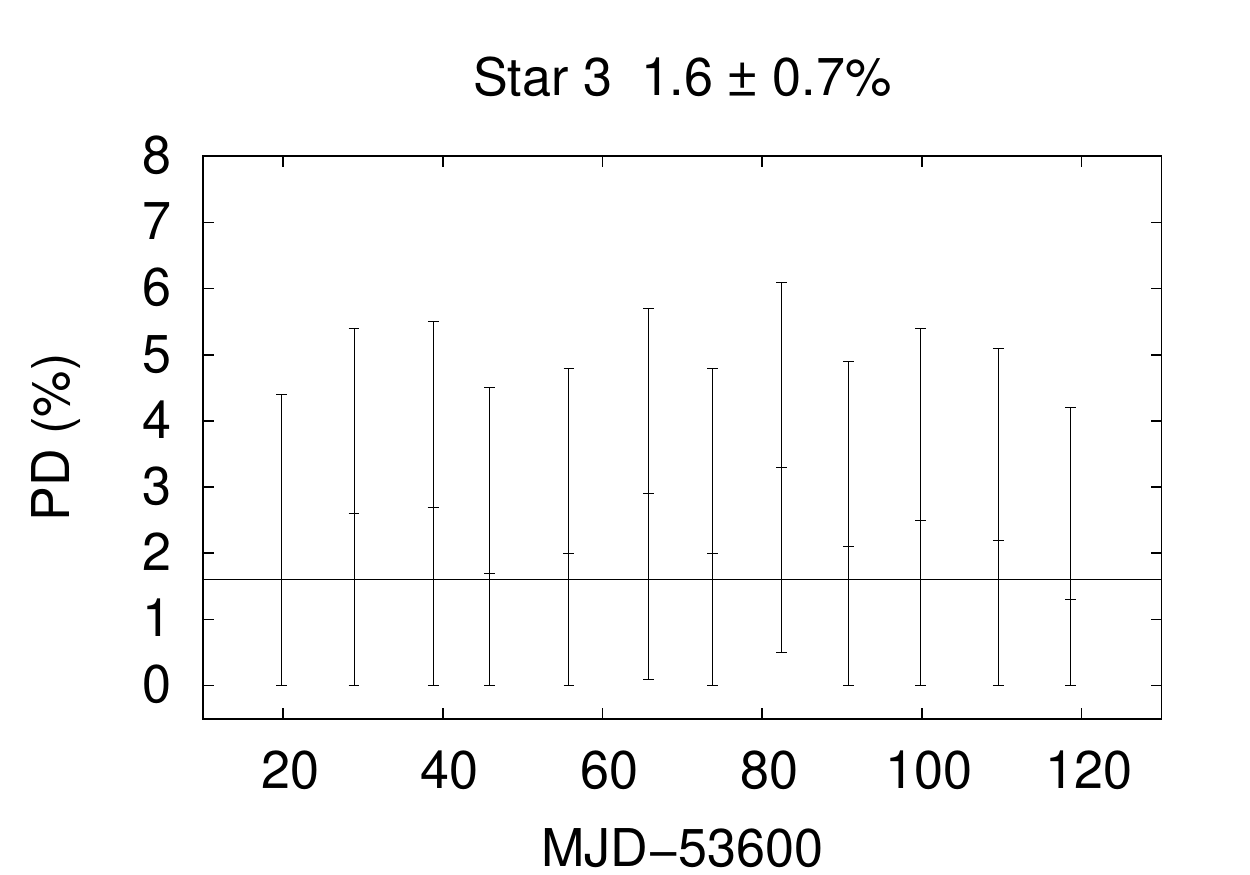}}
\subfloat{\includegraphics[width=45mm]{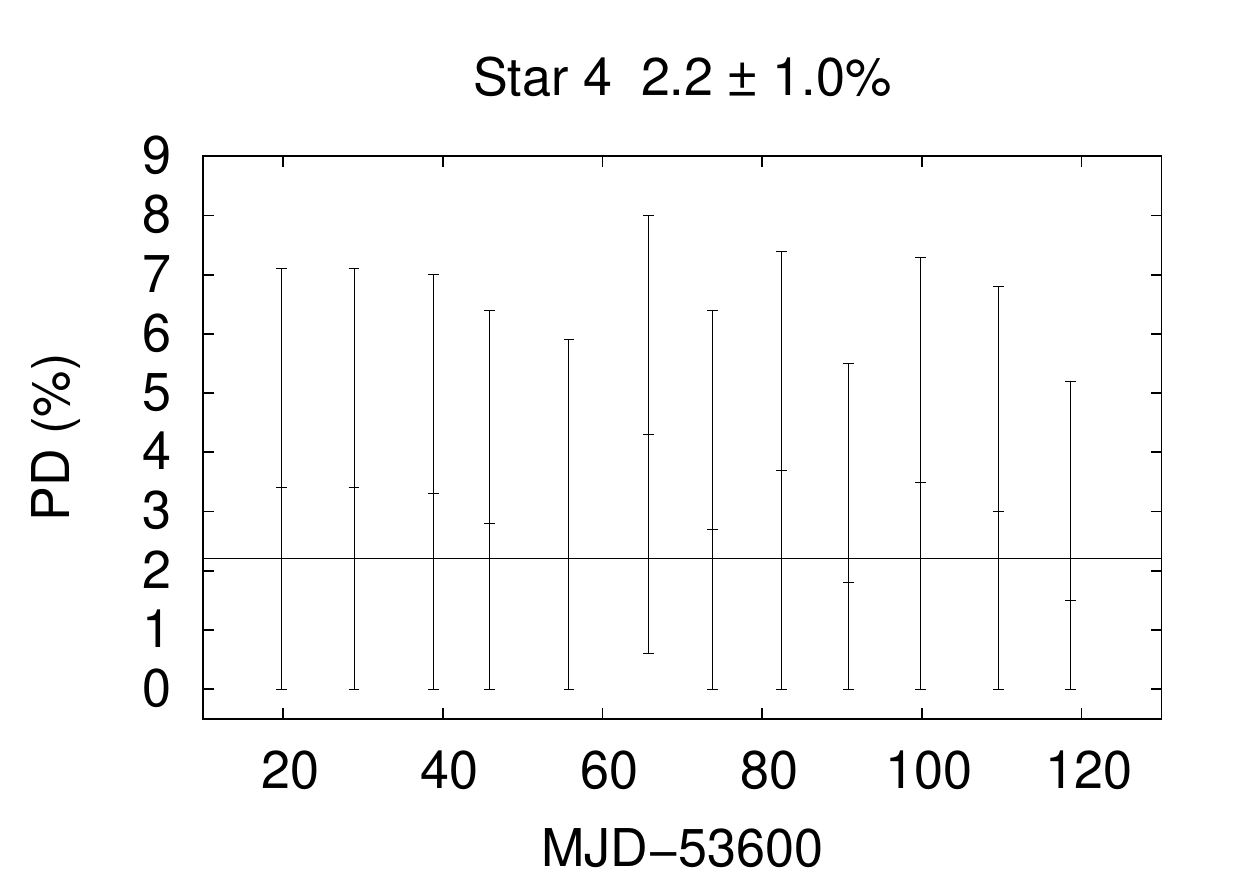}}
\caption{Plots of the degree of linear polarisation (\%) of the sources as a function of time. The solid lines are the weighted mean of the degree of polarisation.}
\label{figure5}
\end{figure*}

\begin{figure*}
\centering
\subfloat{\includegraphics[width=45mm]{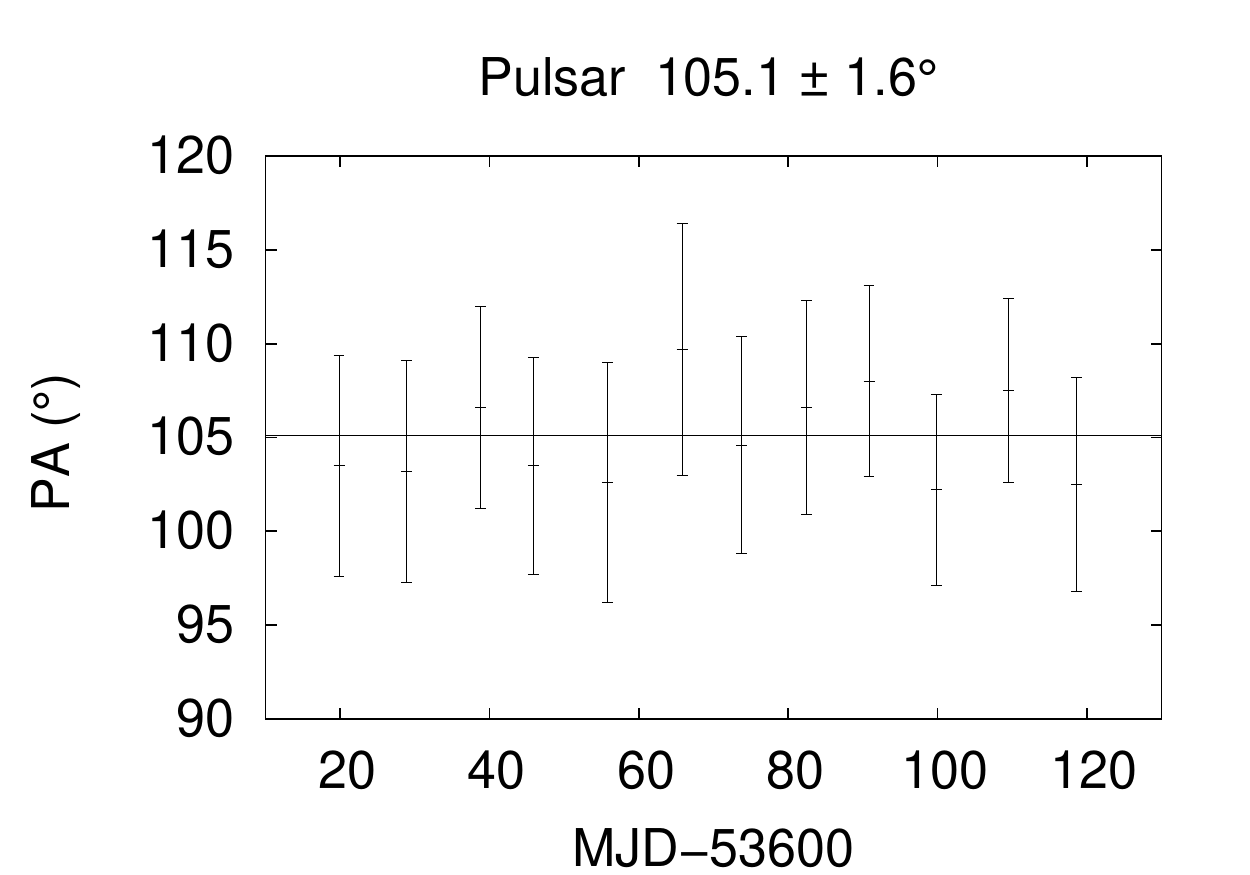}}
\subfloat{\includegraphics[width=45mm]{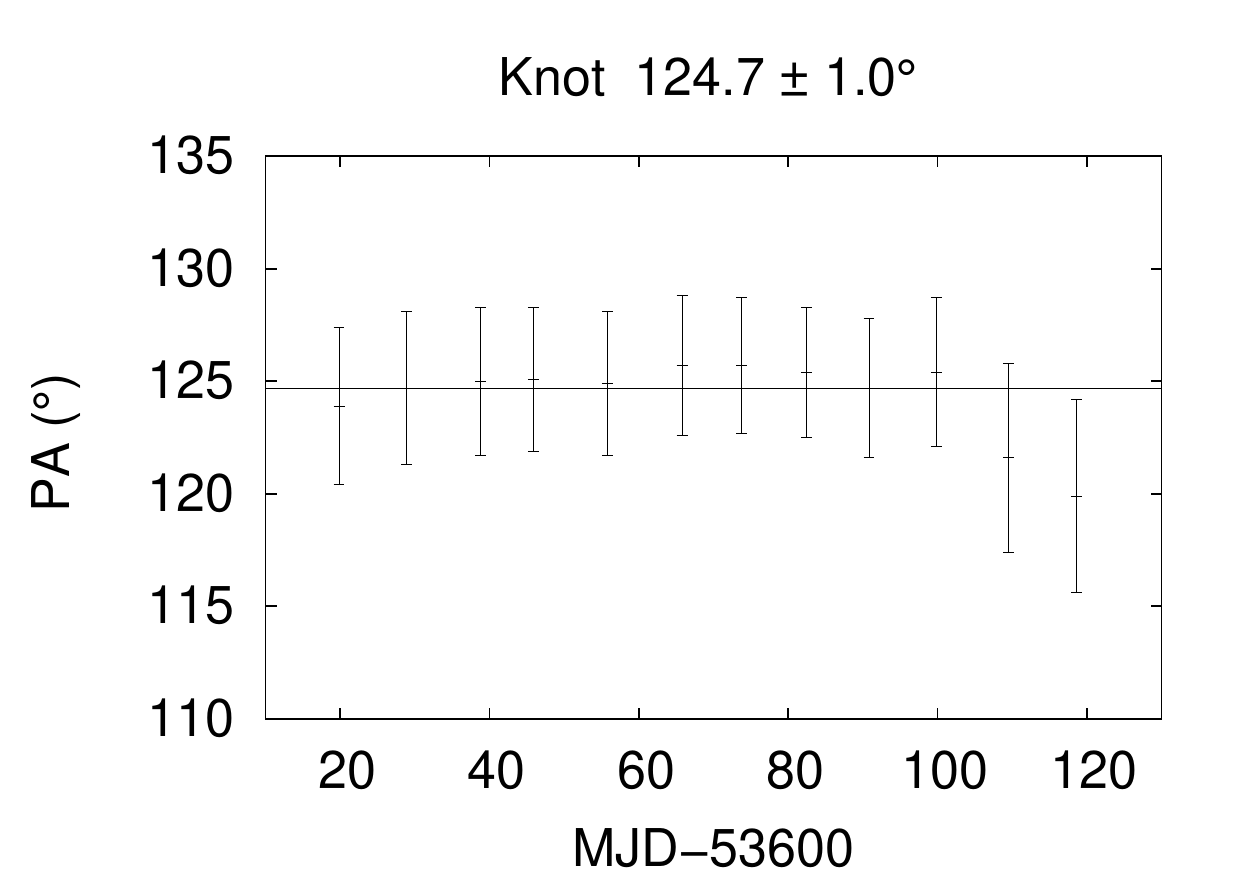}}
\subfloat{\includegraphics[width=45mm]{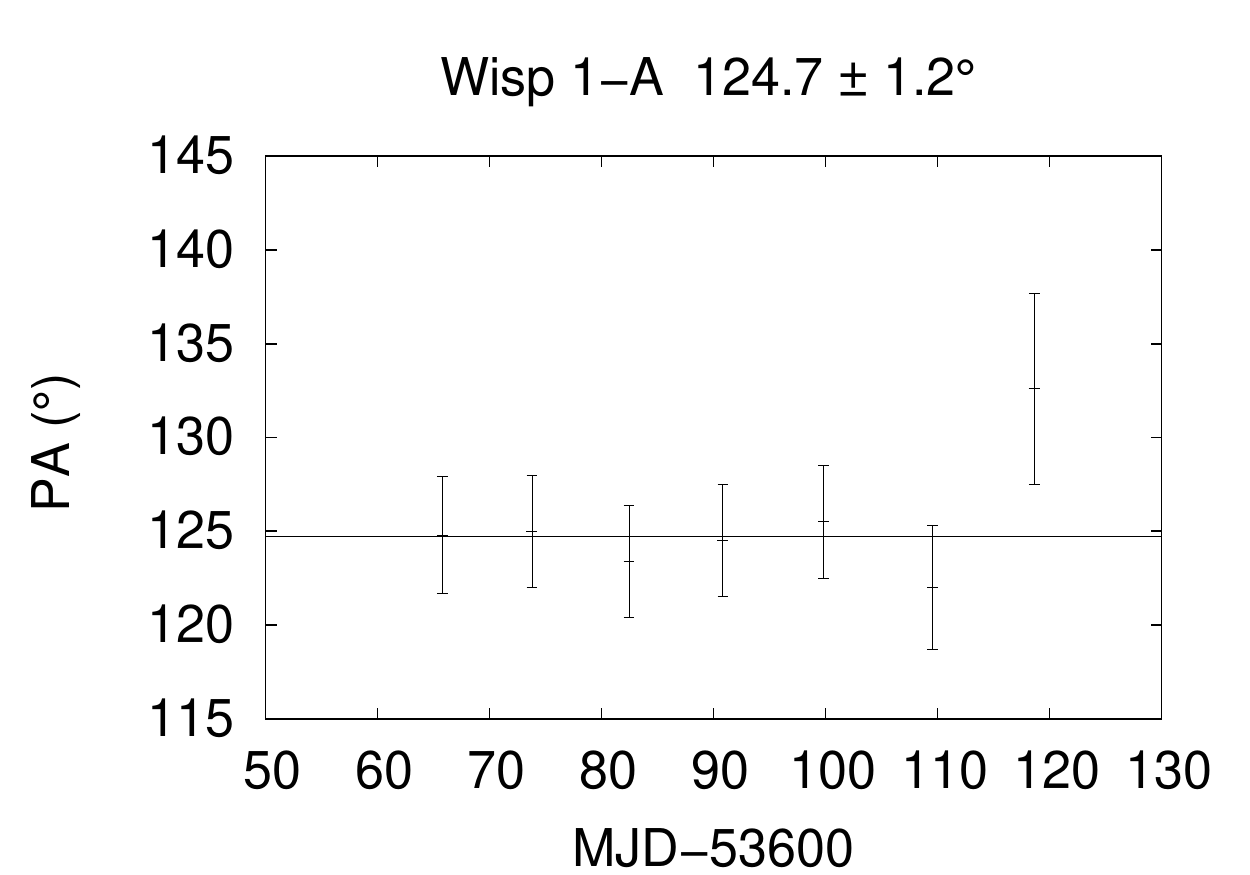}}
\subfloat{\includegraphics[width=45mm]{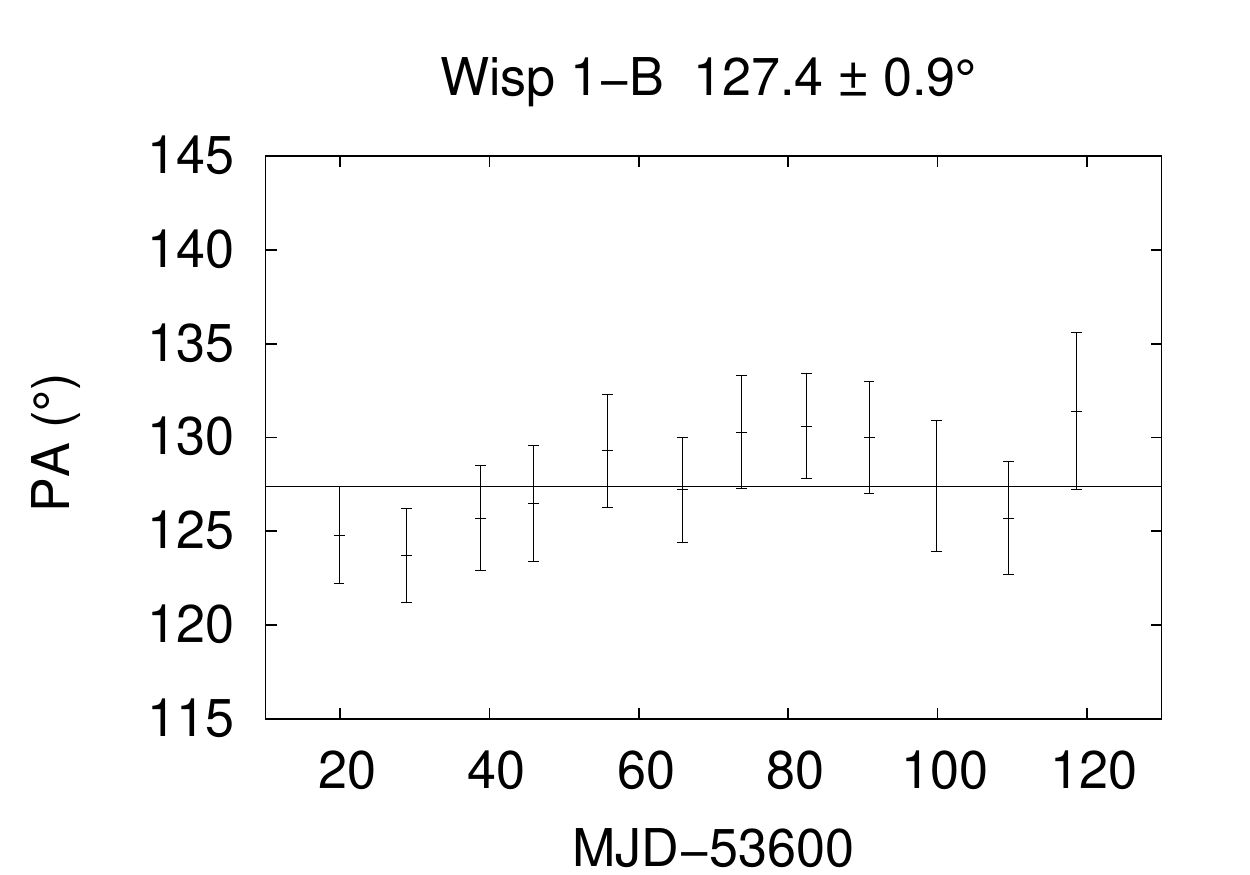}}

\subfloat{\includegraphics[width=45mm]{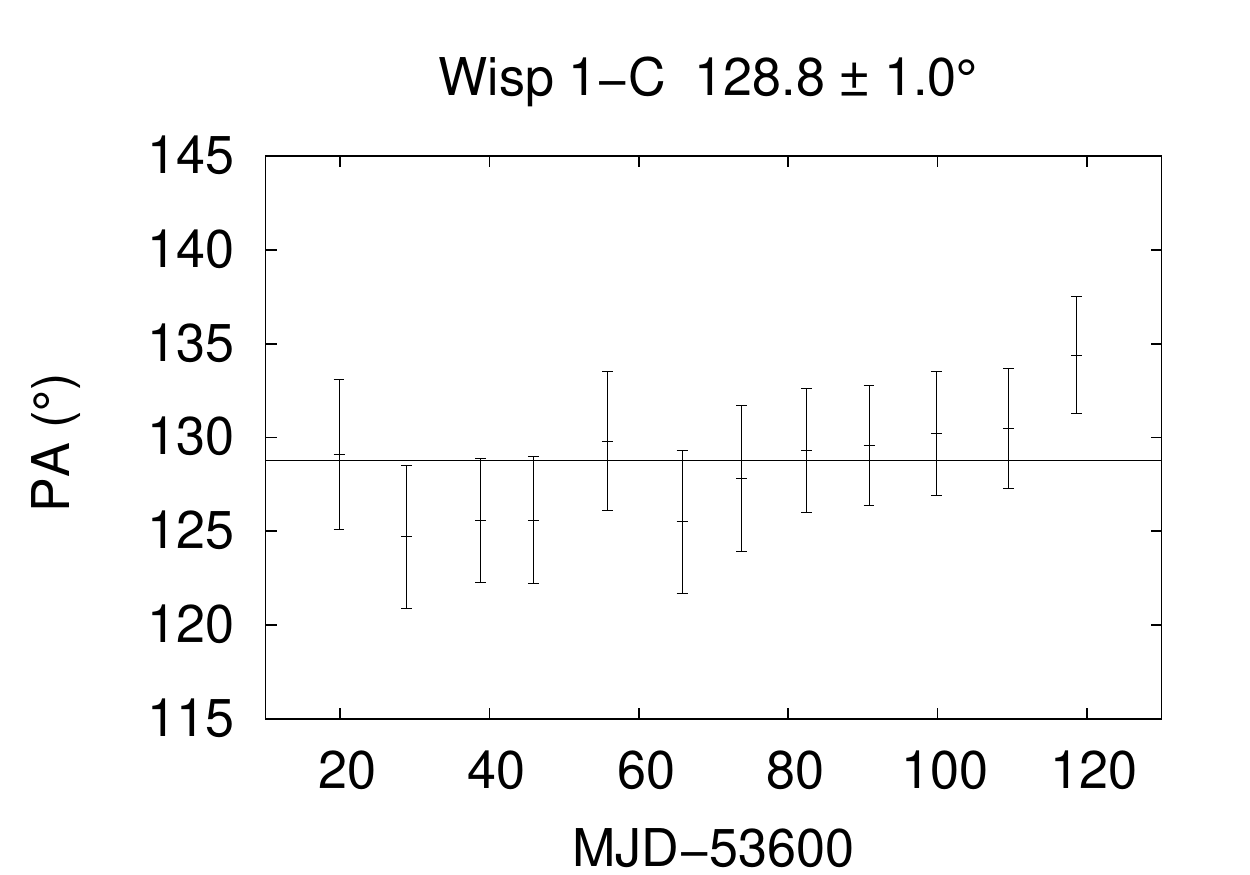}}
\subfloat{\includegraphics[width=45mm]{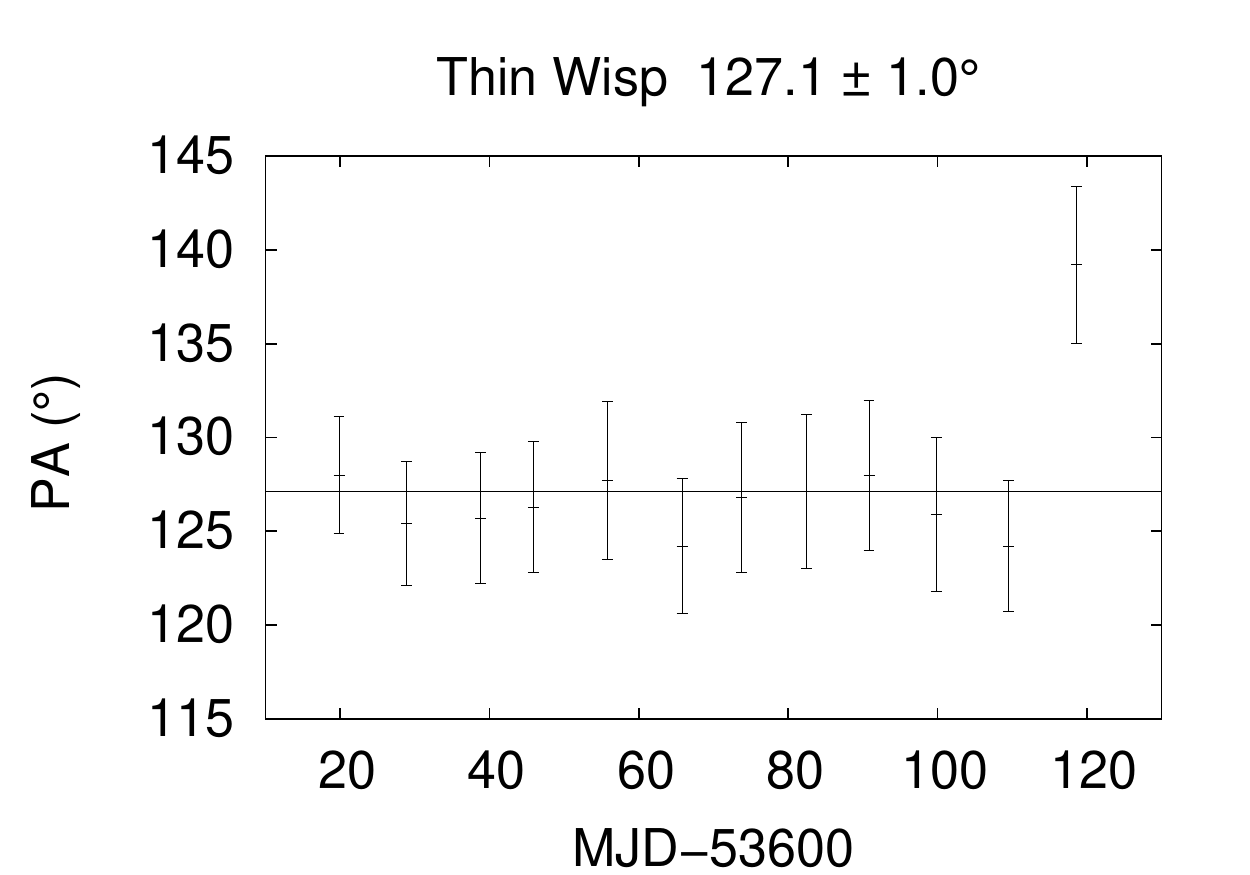}}
\subfloat{\includegraphics[width=45mm]{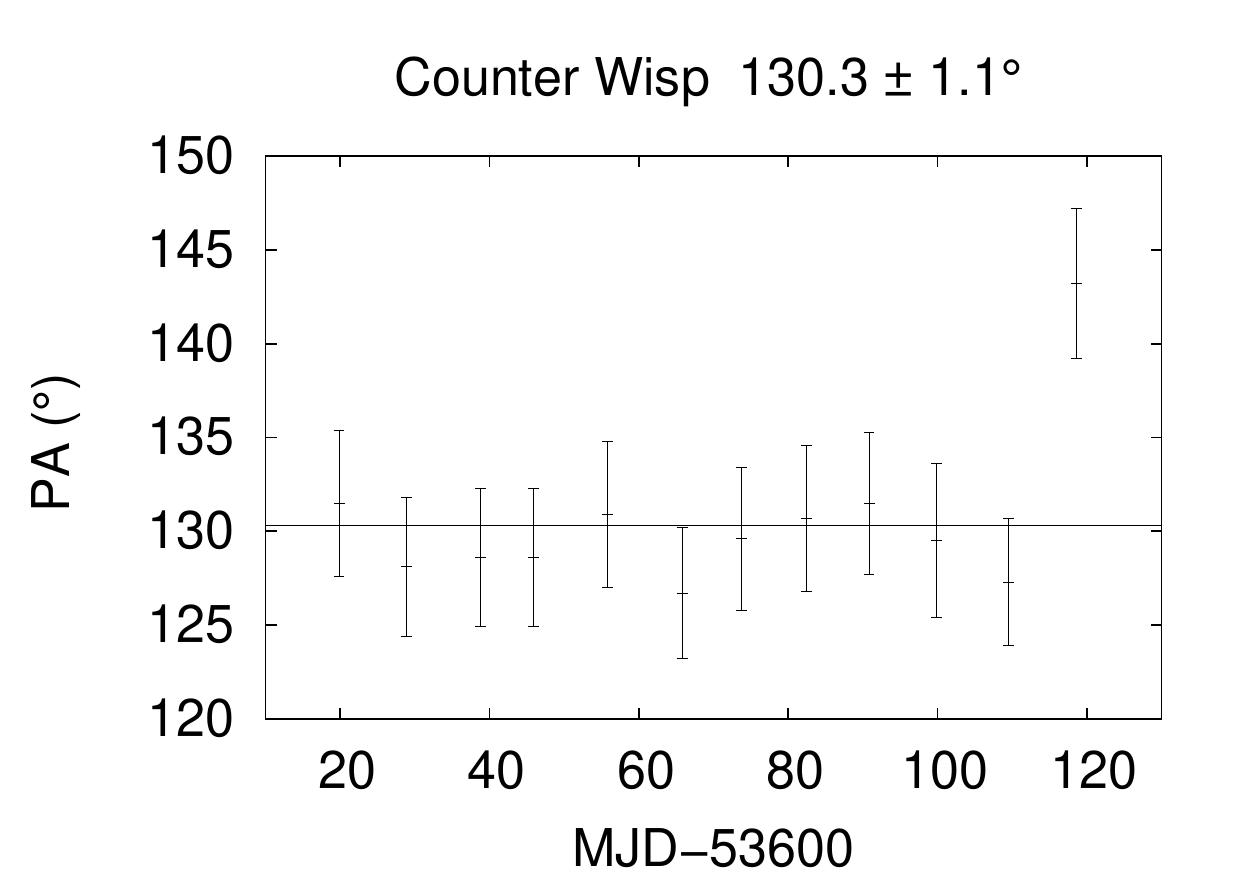}}
\subfloat{\includegraphics[width=45mm]{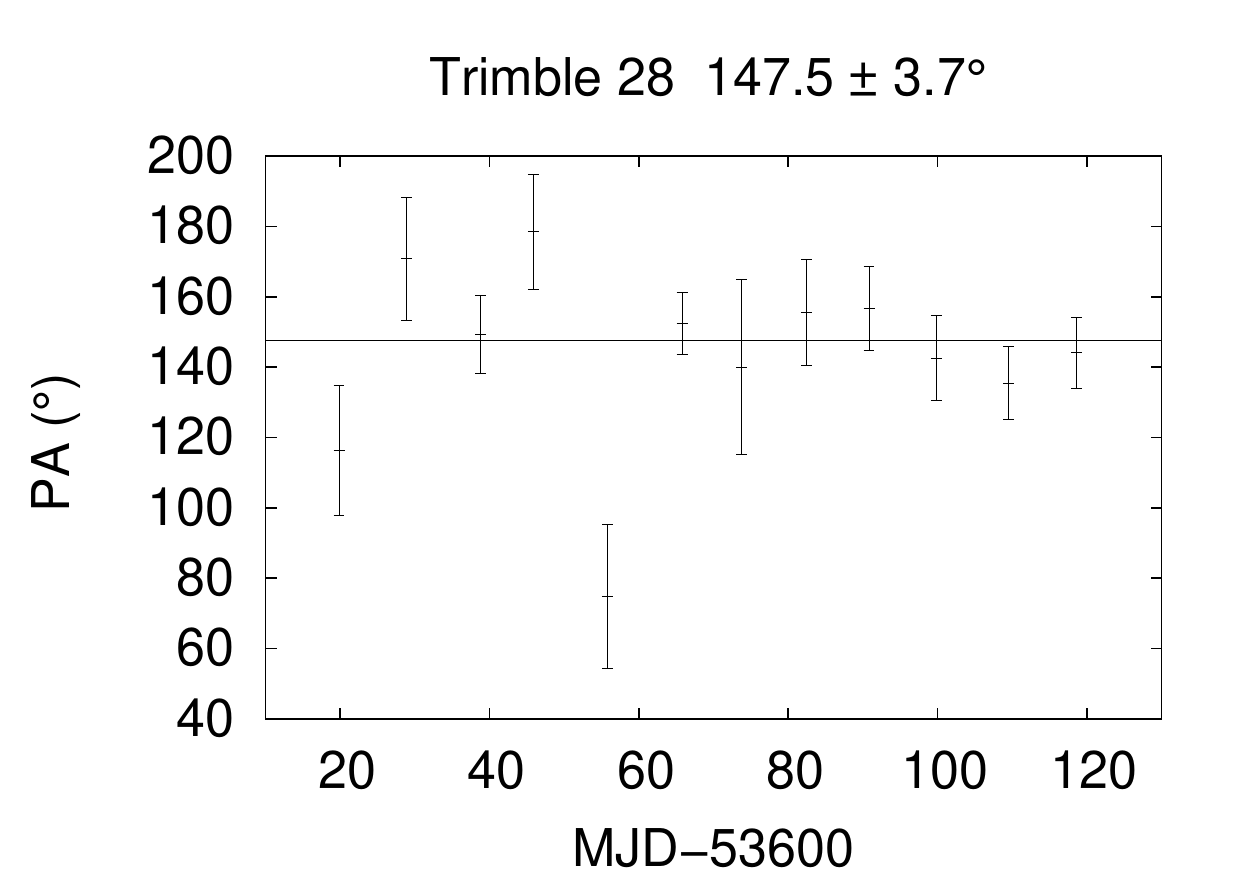}}

\subfloat{\includegraphics[width=45mm]{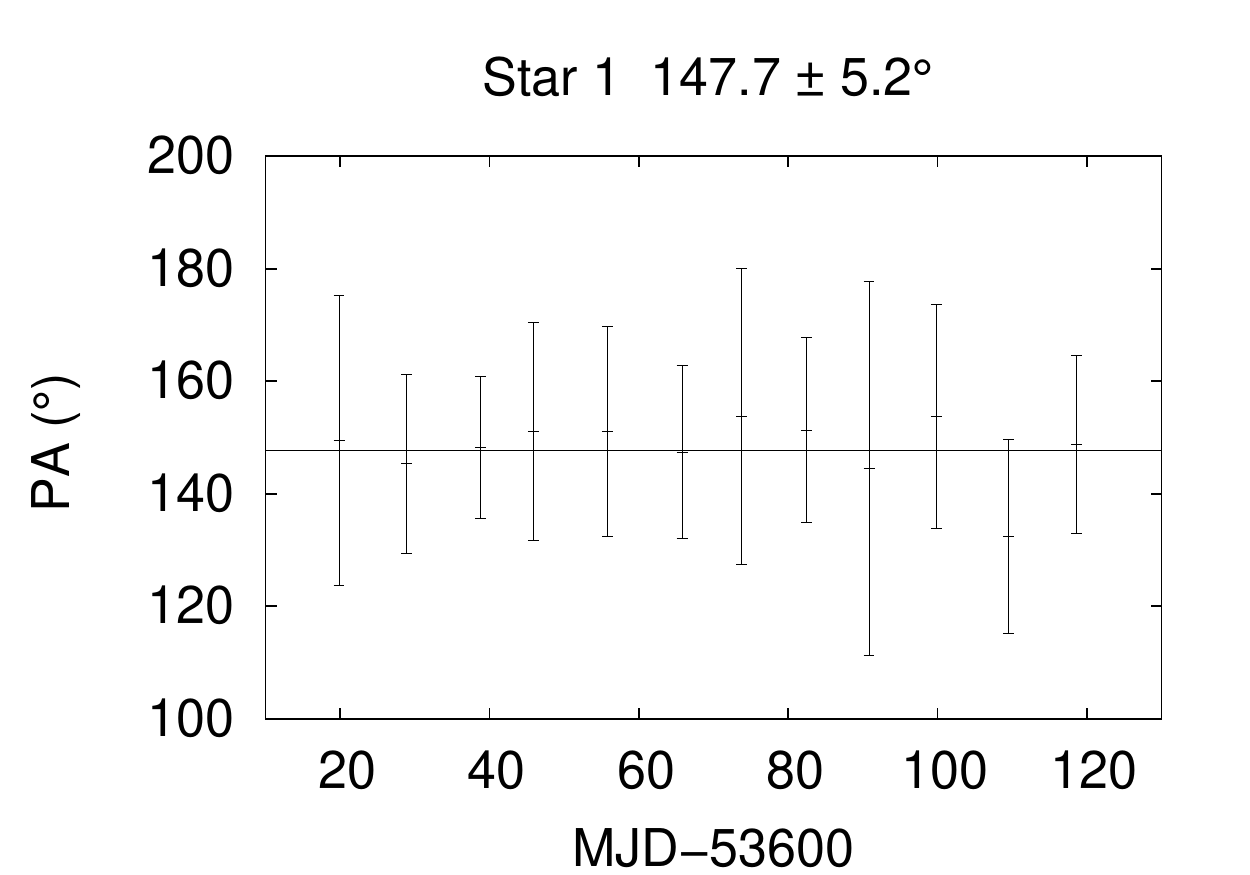}}
\subfloat{\includegraphics[width=45mm]{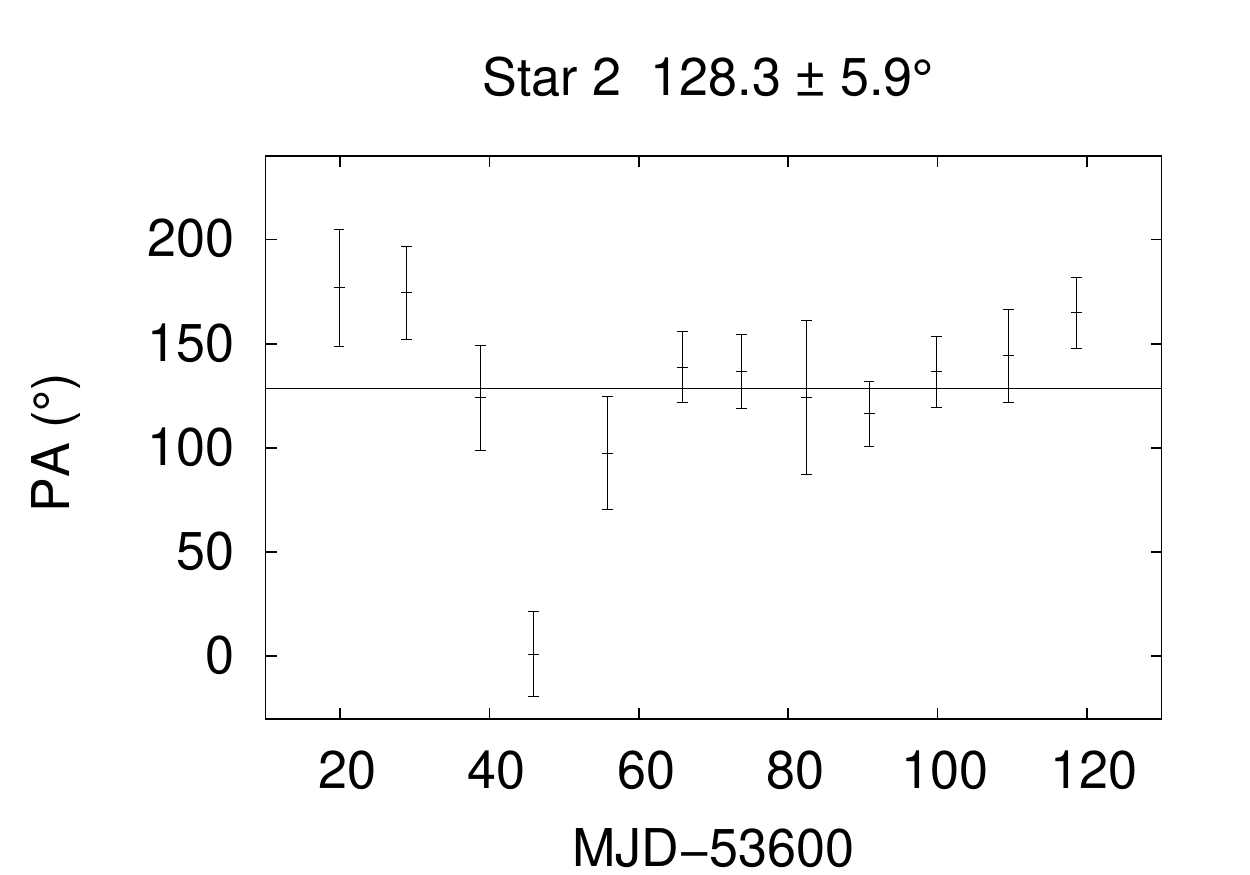}}
\subfloat{\includegraphics[width=45mm]{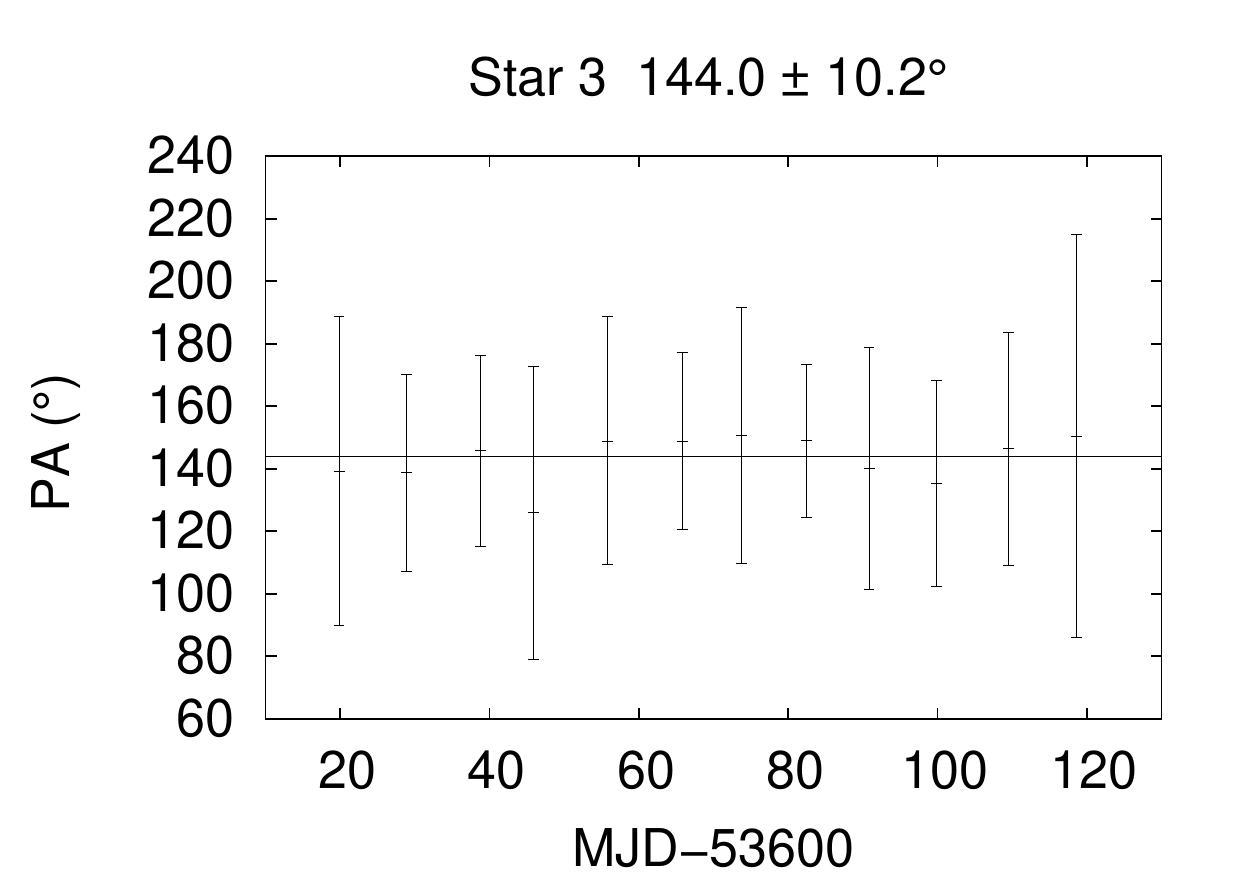}}
\subfloat{\includegraphics[width=45mm]{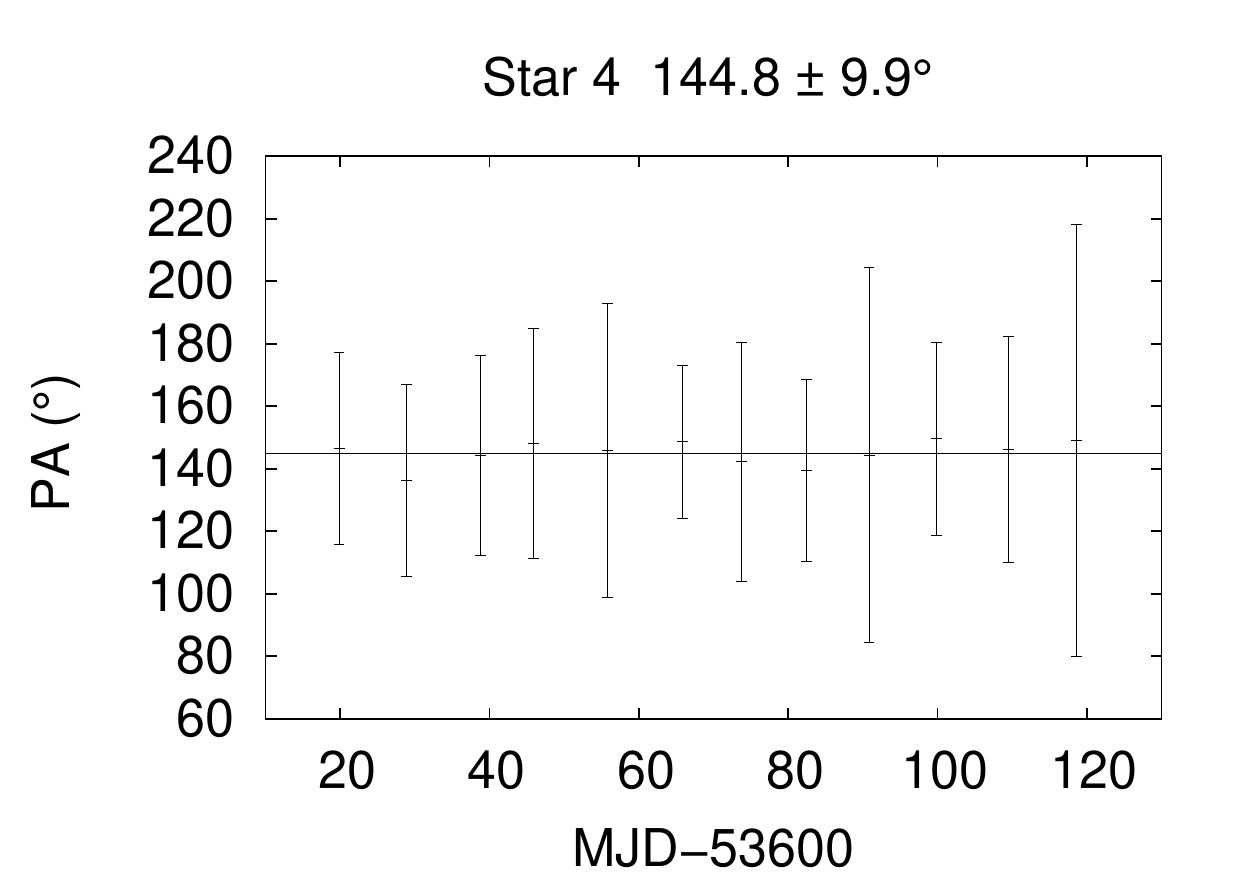}}
\caption{Plots of the polarisation position angle (\degr) of the sources as a function of time. The solid lines are the weighted mean of the position angle.}
\label{figure6}
\end{figure*}


\begin{figure*}
\includegraphics[width=90mm]{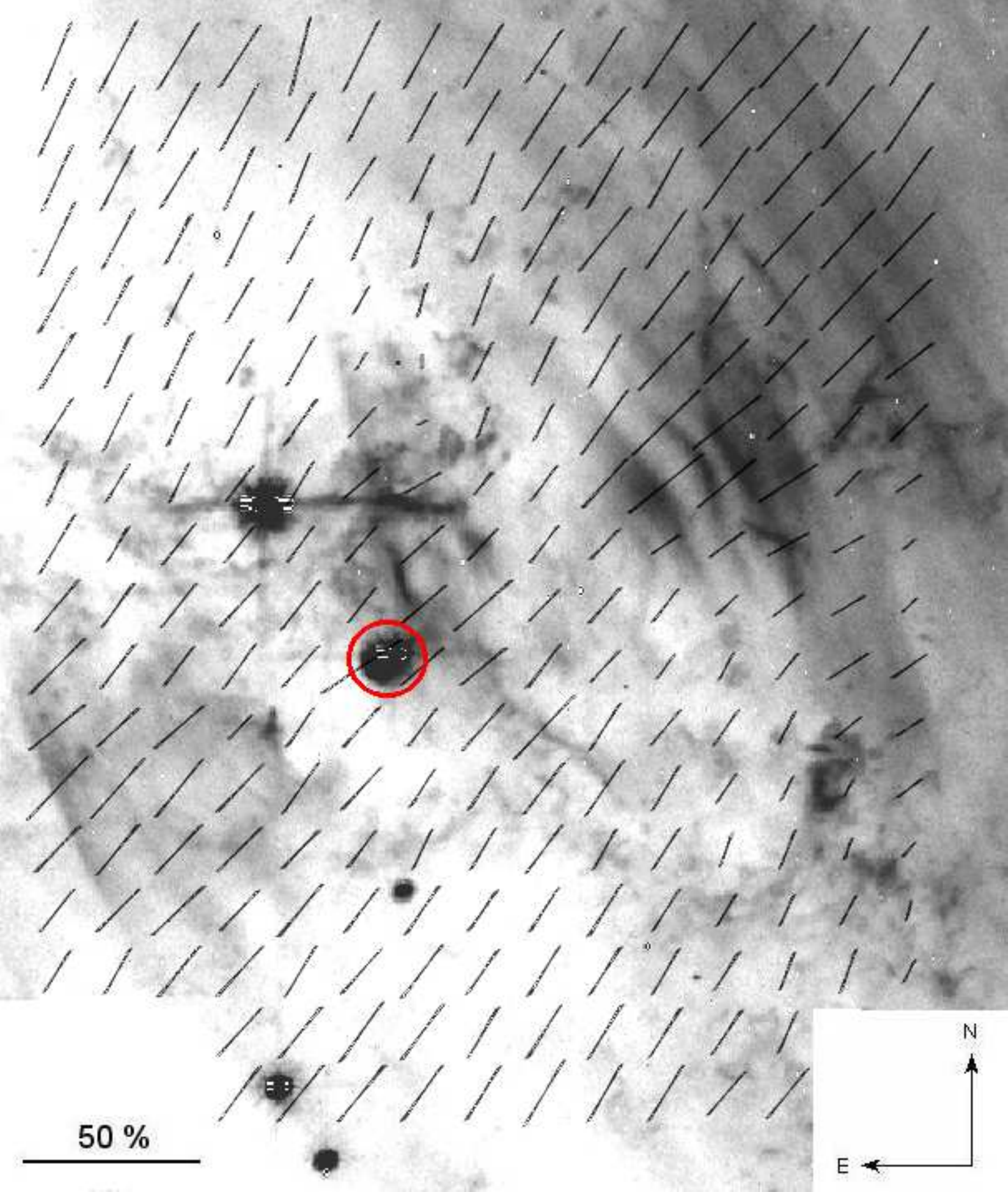}
\caption{Polarisation vector map of the vicinity of the Crab pulsar superimposed on the nebula (2005 Nov 25, FOV $\approx25\arcsec\times33\arcsec$). The location of the pulsar and inner knot is marked by the circle. The legend shows the vector magnitude for 50\% polarisation.}
\label{figure7}
\end{figure*}

\begin{figure*}
\includegraphics[height=100mm, angle=0]{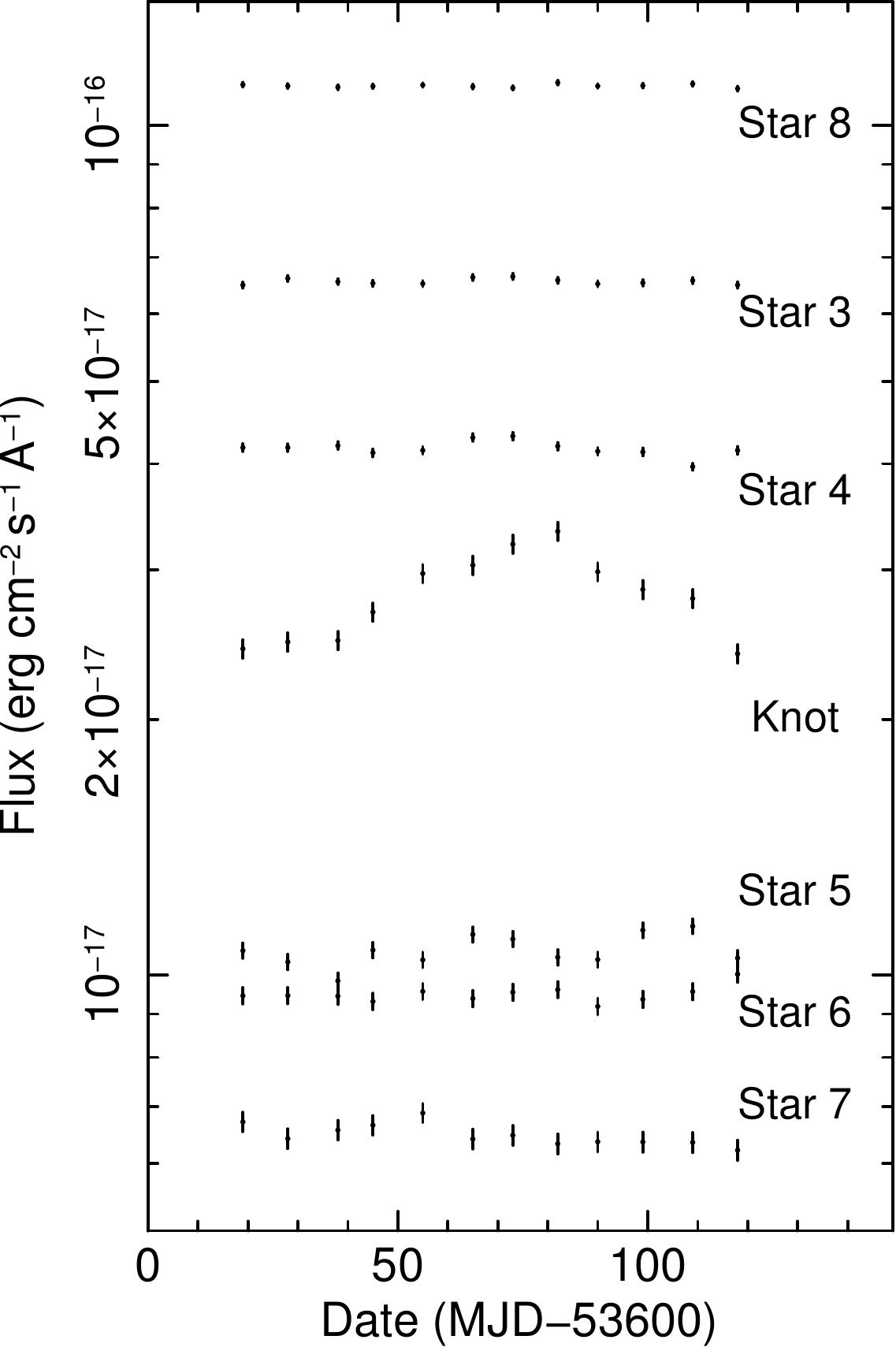}
\caption{Results of photometry on the knot as well as on a sample of field stars. A large flux variability for the knot is apparent, with a $\sim40\%$ brightening on a two-month time scale.}
\label{figure8}
\end{figure*}

\begin{figure*}
\includegraphics[width=65mm]{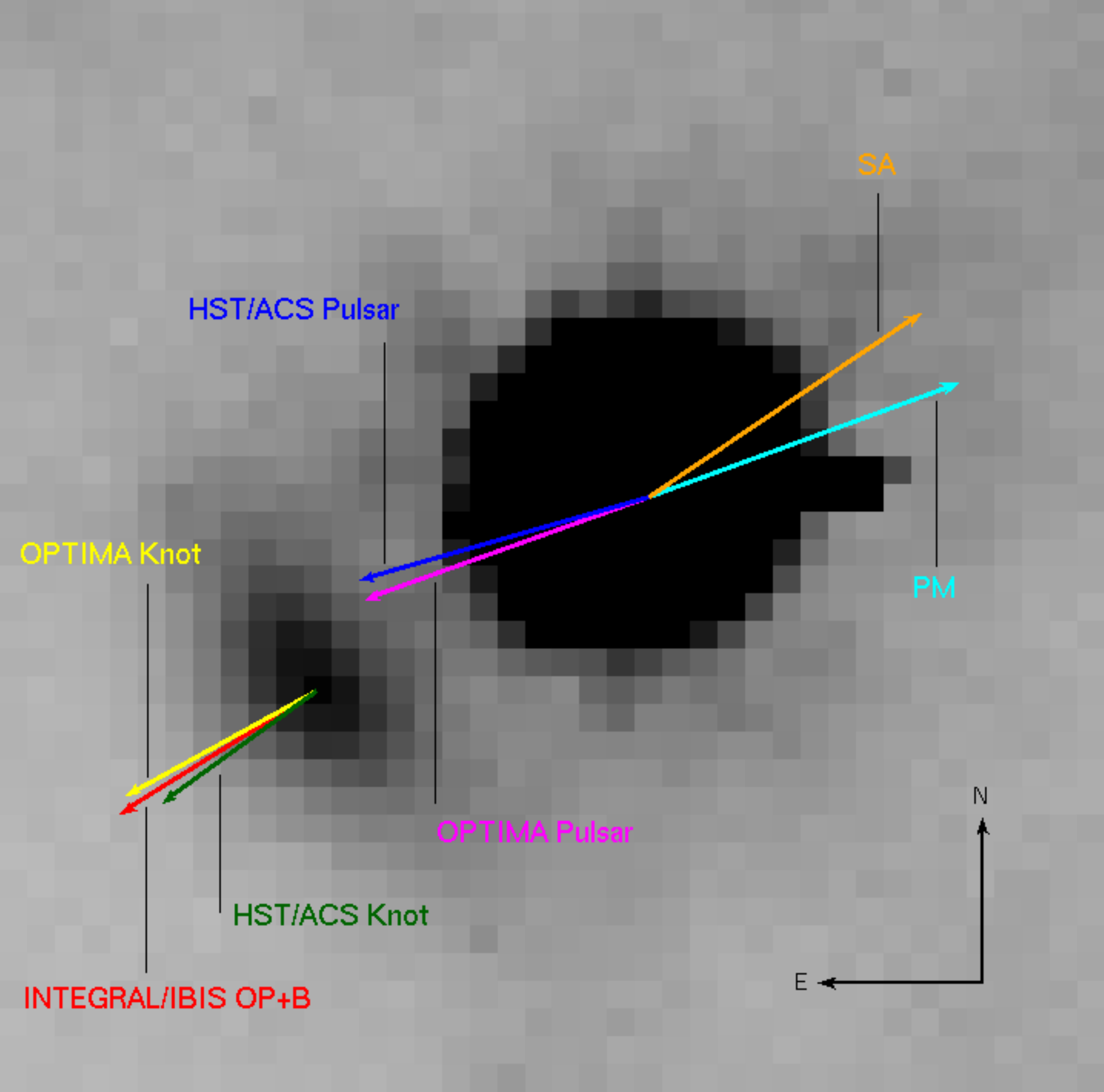}
\caption{The pulsar region with the synchrotron knot located $\approx0\farcs65$ SE of the pulsar (2005 Sep 06, FOV $\approx2\arcsec\times2\arcsec$). The vectors included are as follows: spin-axis vector (SA) ($124\pm0.1\degr$) (Ng \& Romani 2004), proper motion vector (PM) ($110\pm2\pm9\degr$) \citep{Kaplan08}, and the polarisation position angles of the pulsar ($105.1\pm1.6\degr$) and synchrotron knot ($124.7\pm1.0\degr$) from the HST/ACS data. Also, included are the phase-averaged OPTIMA measurements of the polarisation position angles of the synchrotron knot ($119.8\pm0.8\degr$) and pulsar ($109.5\pm0.2\degr$) \citep{Aga}, and the phase-averged INTEGRAL/IBIS measurement of the polarisation position angle during the off-pulse and bridge emission (OP+B) phases ($122.0\pm7.7\degr$) \citep{Forot}.}
\label{figure9}
\end{figure*}

\begin{figure*}
\includegraphics[height=65mm]{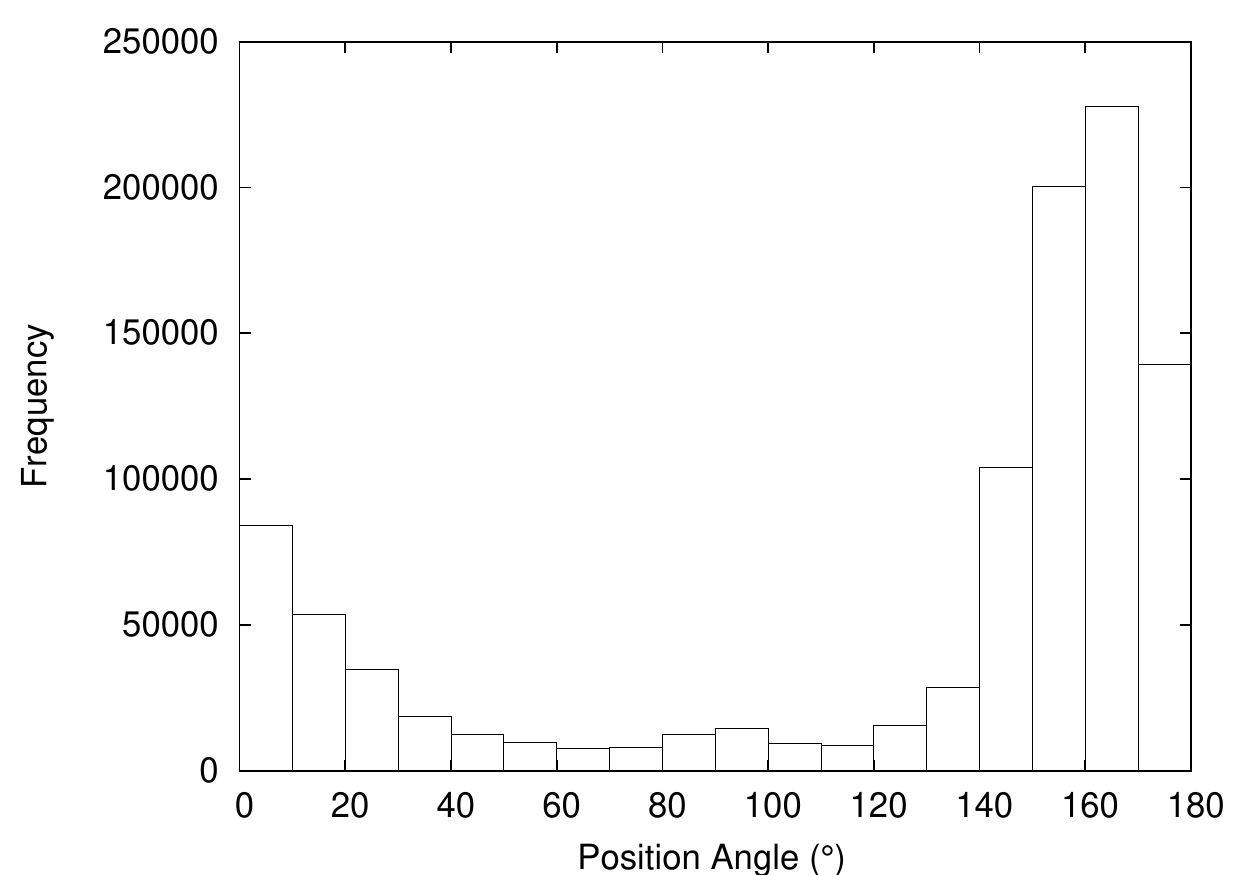}
\caption{Histogram of the polarisation position angles of the entire inner nebula.}
\label{figure8}
\end{figure*}


\section{Discussion}

We have studied the phase-averaged polarisation properties of the Crab pulsar and its nearby synchrotron knot using archival HST/ACS data. We note that the dataset analysed in this paper has previously been used by \citet{Hester08} to examine the morphology and structure of the polarised components of the inner nebula. However, we have produced polarisation vector maps of the inner nebula and measured, for the first time, the degree of linear polarisation and the position angle of the pulsar's integrated pulse beam, and of its nearby synchrotron knot. Furthermore, this work marks the first high-spatial resolution multi-epoch study of the variability of the polarisation of the inner nebula and pulsar.

The results for the Crab pulsar are $\rm P.D.=5.2\pm0.3\%$, and $\rm P.A.=105.1\pm1.6\degr$ (see Table 6). These values are in good agreement with those of \citet{Aga} using the high-time resolution photo-polarimeter OPTIMA\footnote{http://www.mpe.mpg.de/OPTIMA} \citep{Kanbach}, once DC substracted. They measure phase-averaged values of $\rm P.D.=9.8\pm0.1\%$, and $\rm P.A.=109.5\pm0.2\degr$, which is not DC subtracted and includes the emission from the inner knot due to the OPTIMA aperture. They measure values of $\rm P.D.=5.4\%$, and $\rm P.A.=96.4\degr$ after DC subtraction, and it is this later measurement that agrees with our own. The optical polarisation of the Crab pulsar has also been measured by \citet{Wampler} ($\rm P.D.=6.5\pm0.9\%$, $\rm P.A.=107.0\pm6.0\degr$), and \citet{Kristian} ($\rm P.D.=6.8\pm0.5\%$, $\rm P.A.=98.0\pm3.0\degr$). 

We note that the polarisation of the inner knot ($59.0\pm1.9\%$) is a factor of two larger than the off-pulse polarisation of 33\% obtained from OPTIMA observations \citep{Aga} and consistent with the older measurements of \citep{Jones} (70\%) and \citep{Smith} ($47\pm10\%$). This discrepancy is partially due to the uncertainty of determining the phase interval bracketing the minimum of the Crab's lightcurve, hence the contribution of the DC component \citep[see Fig. 5 of][]{Aga}. It could also be partially due to uncertainties in the estimate of the sky background in the OPTIMA data and/or the contribution from the sky and pulsar \lq\lq off-pulse\rq\rq\ flux. \citet{Golden} give the unpulsed pulsar flux to be 0.02 mJy compared to 0.03 mJy from the knot (this work). We estimate the contribution from the sky for OPTIMA data to be equivalent to 0.04 mJy based on the pupil size of 2\farcs35 and a 21 magnitude/arcsec$^{2}$ sky background. This would be sufficient to explain the difference. Two-dimensional phase-resolved polarisation observations will allow us to better quantify the knot contribution to the DC component.

X-ray observations of the nebula taken by Chandra \citep{Weisskopf2000}, reveal a torus with bipolar jets emanating outwards from SE and NW of the pulsar. \citet{Ng06} found that the axis of symmetry of the jet is rougly aligned with the pulsar's proper motion vector. The Crab torus, bisecting the synchrotron wisps, can be traced back to the knot of synchrotron emission seen $\approx0\farcs65$ SE of the pulsar. Our measurement of the polarisation PA of the synchrotron knot, $\rm PA=124.7\pm1.0\degr$, agrees with the Crab torus $\rm PA=126.31\pm0.03\degr$ \citep{Ng04}. We also found evidence for an apparent alignment between the pulsar polarisation PA (105.1$\pm1.6\degr$) and proper motion vector (\citealt{Kaplan08}; 110$\pm2\pm9\degr$) (see Figure 9). \citet{Mignani} have found the same scenario for the Vela pulsar. Those authors found an apparent alignment between the polarisation position angle of the pulsar, the axis of symmetry of the X-ray arcs and jets (Chandra; \citet{Pavlov}, \citet{Helfand}), and the pulsar’s proper motion vector. This suggests that the \lq\lq kick\rq\rq\ given to neutron stars at birth is directed along the rotation axis \citep{Lai}. The alternative view is that the apparent alignment is an effect of projection onto the sky plane, and that there is no physical jet along the axis of rotation \citep{RD}. More concrete measurements of the optical polarisation of pulsars will yield the needed observational restraints on these hypotheses.\\
\indent As mentioned previously, the polarisation of the wisps was also studied. Our photometry accounts for the outward motion of the wisps. From analysis of the wisps in each epoch, we find that the wisps show variation in both location and brightness on time scales of a few weeks. We found that all of these wisps have similar values of degree of polarisation ($\sim 40\%$) and position angles equal to that of the synchrotron knot ($\sim 125\degr$). Hence, as with the synchrotron knot, they are aligned with the spin-axis of the pulsar. Also, Wisps 1-A is not visible in the frames from 2005 Spetember to 2005 October 12 inclusive, and may be merged with Wisp 1-B during this period. Examining the polarisation vectors maps, one can see that the position angles of the wisps are different to those of the rest of the nebula, where the position angles are aligned NS (Figures 2 and 7).  Figure 10 is a histogram of the distribution of the polarisation position angles of the inner nebula. The position angles were extracted from the values in the polarisation map. From this histogram we see that the polarisation position angle of the pulsar environment ($\sim 125\degr$) is away from the peak of the nebula distribution ($\sim 165\degr$). This means that the polarisation properties of the structures close to the pulsar are different from those of the rest of the inner nebula. \\
\indent As discussed earlier, using a $\chi^{2}$ goodness-of-fit, we found no significant variation (at the 95\% confidence level) in the polarisation of the pulsar, knot, and wisps over a 3 month period. The knot is variable in flux but fairly constant in polarisation. This variation in flux may be explained in terms of an increased plasma density in the vicinity of the knot. Whereas the wisps have constant flux and constant polarisation over this period of time. This would suggest that the magnetic fields within the nebula are uniform over time. However, more detailed follow-up observations will be needed to determine if there is any longer term variation.


\section{Conclusions}

We have studied the phase-averaged polarisation properties of the Crab pulsar and its nearby synchrotron knot using archival HST/ACS data. This marks the first high-spatial resolution multi-epoch study of the polarisation of the inner nebula and pulsar. We found an apparent alignment between the polarisation position angle of the pulsar and the pulsar's proper motion vector. We confirm that the inner knot is responsible for the highly polarised off-pulse emission seen in observations in the optical. We found that the inner knot is variable in position, and brightness over the period of these observations. These are the first quantified measurements of such a variation. We note that we found evidence of a possible variation of the knot polarisation (at $2 \sigma$) which is due neither to a known systematic effect nor to the spike contribution. Future observations will help to address this point. We have also measured the polarisation of the wisps in the inner nebula, and found no significant variation in their polarisation over this 3 month period of observations.\\
\indent Polarisation measurements give an unique insight into the geometry of the pulsar emission regions. More multi-wavelength polarisation observations of pulsars, both phase-averaged and phase-resolved, with instruments such as HST/ACS and GASP (optical), and INTEGRAL/IBIS (gamma-ray), will help to provide the much needed data to constrain the theoretical models.\\
\indent For example, \citet {McDonald} have developed an inverse mapping approach for determining the emission height of the optical photons from pulsars. It uses the optical Stokes parameters to determine the most likely geometry for emission, including: magnetic field inclination angle ($\alpha$), the observers line of sight angle ($\chi$), and emission height.


\section*{Acknowledgments} 

All of the data presented in this paper were obtained from the Mikulski Archive for Space Telescopes (MAST). STScI is operated by the Association of Universities for Research in Astronomy, Inc., under NASA contract NAS5-26555. Support for MAST for non-HST data is provided by the NASA Office of Space Science via grant NNX09AF08G and by other grants and contracts. We thank Jeremy Walsh, ESO, for the use of his polarimetry software IMPOL to produce the polarisation maps. PM is grateful for his funding from the Irish Research Council (IRC). RPM thanks the European Commission Seventh Framework Programme (FP7/2007-2013) for their support under grant agreement n.267251. AS\l{} acknowledges support from the Foundation for Polish Science grant FNP HOM/2009/11B, as well as from the FP7 Marie Curie European Reintegration Grant (PERG05-GA-2009-249168). This work was in part supported under the FP7 Opticon European Network for High Time Resolution Astrophysics (HTRA) project.



\label{lastpage}

\end{document}